\documentclass[fleqn,usenatbib]{mnras}

\usepackage{newtxtext,newtxmath}

\usepackage[T1]{fontenc}
\usepackage{ae,aecompl}

\DeclareRobustCommand{\VAN}[3]{#2}
\let\VANthebibliography\thebibliography
\def\thebibliography{\DeclareRobustCommand{\VAN}[3]{##3}\VANthebibliography}

\usepackage{graphicx}	
\usepackage{amsmath}	

\title[Precise light curves through direct imaging]{Chasing the storm: Investigating the application of high-contrast imaging techniques in producing precise exoplanet light curves} 

\author[B. J. Sutlieff et al.]{Ben J. Sutlieff,$^{1,2,3,4}$\thanks{E-mail: ben.sutlieff@roe.ac.uk}
David S. Doelman,$^{4,5}$
Jayne L. Birkby,$^{6}$
Matthew A. Kenworthy,$^{4}$
Jordan M. Stone,$^{7}$
\newauthor Frans Snik,$^{4}$
Steve Ertel,$^{8,9}$
Beth A. Biller,$^{1,2}$
Charles E. Woodward,$^{10}$
Andrew J. Skemer,$^{11}$
\newauthor Jarron M. Leisenring,$^{8}$
Alexander J. Bohn,$^{4}$
and Luke T. Parker$^{6}$
\\
$^{1}$Institute for Astronomy, University of Edinburgh, Royal Observatory, Blackford Hill, Edinburgh, EH9 3HJ, UK\\
$^{2}$Centre for Exoplanet Science, University of Edinburgh, Edinburgh, EH9 3HJ, UK\\
$^{3}$Anton Pannekoek Institute for Astronomy, University of Amsterdam, Science Park 904, 1098 XH Amsterdam, The Netherlands\\
$^{4}$Leiden Observatory, Leiden University, P.O. Box 9513, 2300 RA Leiden, The Netherlands\\
$^{5}$SRON Netherlands Institute for Space Research, Niels Bohrweg 4, 2333 CA, Leiden, The Netherlands\\
$^{6}$Astrophysics, University of Oxford, Denys Wilkinson Building, Keble Road, Oxford, OX1 3RH, United Kingdom\\
$^{7}$Naval Research Laboratory, Remote Sensing Division, 4555 Overlook Ave. SW, Washington, DC 20375, USA\\
$^{8}$Department of Astronomy and Steward Observatory, University of Arizona, 933 N. Cherry Ave., Tucson, AZ 85721, USA\\
$^{9}$Large Binocular Telescope Observatory, University of Arizona, 933 North Cherry Avenue, Tucson, AZ 85721, USA\\
$^{10}$Minnesota Institute for Astrophysics, University of Minnesota, 116 Church Street SE, Minneapolis, MN 55455, USA\\
$^{11}$Department of Astronomy and Astrophysics, University of California, Santa Cruz, 1156 High St, Santa Cruz, CA 95064, USA
}

\date{Accepted 2025 October 31. Received 2025 October 30; in original form 2025 April 29}

\pubyear{2025}

\begin{document}
\label{firstpage}
\pagerange{\pageref{firstpage}--\pageref{lastpage}}
\maketitle

\begin{abstract}
Substellar companions such as exoplanets and brown dwarfs exhibit changes in brightness arising from top-of-atmosphere inhomogeneities, providing insights into their atmospheric structure and dynamics. This variability can be measured in the light curves of high-contrast companions from the ground by combining differential spectrophotometric monitoring techniques with high-contrast imaging. However, ground-based observations are sensitive to the effects of turbulence in Earth's atmosphere, and while adaptive optics (AO) systems and bespoke data processing techniques help to mitigate these, residual systematics can limit photometric precision. Here, we inject artificial companions to data obtained with an AO system and a vector Apodizing Phase Plate coronagraph to test the level to which telluric and other systematics contaminate such light curves, and thus how well their known variability signals can be recovered. We find that varying companions are distinguishable from non-varying companions, but that variability amplitudes and periods cannot be accurately recovered when observations cover only a small number of periods. Residual systematics remain above the photon noise in the light curves but have not yet reached a noise floor. We also simulate observations to assess how specific systematic sources, such as non-common path aberrations and AO residuals, can impact aperture photometry as a companion moves through pupil-stabilised data. We show that only the lowest-order aberrations are likely to affect flux measurements, but that thermal background noise is the dominant source of scatter in raw companion photometry. Predictive control and focal-plane wavefront sensing techniques will help to further reduce systematics in data of this type.
\end{abstract}




\section{Introduction}\label{p4_ch5_intro}
Periodic variations in the brightness of rotating exoplanets and brown dwarfs provide a unique avenue to explore their atmospheric structures and how they change over time. Such variations can arise from a range of sources, including inhomogeneous cloud cover, magnetic spots, aurorae, and temperature fluctuations caused by radiative convection, and can have different magnitudes and phases at different wavelengths \citep[e.g.][]{2012MNRAS.427.3358G, 2013ApJ...768..121A, 2017Sci...357..683A, 2014ApJ...793...75R, 2015Natur.523..568H, 2016ApJ...817L..19T, 2020A&A...643A..23T, 2016ApJ...826....8Y, 2019ApJ...874..111T, 2023ApJ...944..138V, 2024ApJ...965...83M, 2025ApJ...981L..22M}. Thus, the variability properties of substellar objects yield valuable information about the underlying physical processes that govern their atmospheres. Time-resolved photometric monitoring has now identified variability in the light curves of many substellar companions and isolated objects \citep[e.g.][]{2015ApJ...799..154M, 2016ApJ...823..152C, 2018AJ....155...95B, 2024MNRAS.532.2207B, 2019ApJ...875L..15M, 2021AJ....162..179M, 2019ApJ...883..181M, 2023MNRAS.521..952M, 2022ApJ...924...68V, 2020AJ....159..125L, 2020ApJ...903...15L, 2021AJ....161..224T, 2016ApJ...818..176Z, 2022AJ....164..239Z, 2024MNRAS.527.6624L}.

Large diameter ground-based telescopes with high-order adaptive optics (AO) systems feeding coronagraphic imagers allow us to resolve substellar companions at close angular separations that are otherwise inaccessible to space-based observatories with smaller mirrors. However, achieving the photometric precision required to measure the variability of these companions can be challenging, as ground-based observations inherently suffer from systematics caused by turbulence in Earth's atmosphere. High-contrast imaging data are often limited by quasi-static speckles of residual starlight at the smallest separations, as well as the wind-driven halo effect that arises when atmospheric turbulence varies faster than the AO system can correct for it \citep[e.g.][]{2007ApJ...654..633H, 2018A&A...620L..10C, 2020A&A...638A..98C, 2019JATIS...5d9003M, 2021PASP..133j4504M}. Non-common path aberrations (NCPAs), introduced by differences in the optical paths that lead to the wavefront sensor of the AO system and the detector, further give rise to changes in the shapes and sizes of the Point Spread Function (PSF) of the target \citep[e.g.][]{2007JOSAA..24.2334S, 2013A&A...555A..94N, 2014SPIE.9148E..5HN, 2018SPIE10703E..1TM, 2019A&A...632A..48B, 2020SPIE11447E..2LM, 2019A&A...629A..11V, 2022A&A...660A.140V, 2022A&A...659A.170S}. Although extreme AO systems and optimised data processing strategies help to significantly reduce these effects, remaining systematics can produce non-astrophysical variability in the light curves of companions.

For observations of isolated objects that are observed without a coronagraph, non-variable comparison stars are often used as simultaneous photometric references to divide out this systematic variability from the photometry of the target \citep[e.g.][]{2009ApJ...701.1534A, 2014A&A...566A.111W, 2015ApJ...813L..23B, 2017AJ....154..138N, 2019MNRAS.483..480V}. However, there are often no comparison stars available in the small fields of view of the coronagraphic imagers used to observe faint companions, and the companion's host star is typically obscured by the coronagraph itself \citep[e.g.][]{2018SPIE10698E..2SR, 2025ARA&A..63..179K}. Some studies have successfully used satellite spots produced by AO systems as photometric references to obtain upper limits of companion variability at the $\gtrsim$10\% level, but also found that these spots themselves vary, preventing deeper sensitivities from being reached \citep{2016ApJ...820...40A, 2021MNRAS.503..743B, 2022AJ....164..143W}.

Nonetheless, differential light curves of close-separation companions can be produced using the technique of differential spectrophotometry when combined with a vector Apodizing Phase Plate (vAPP) coronagraph \citep[][]{2023MNRAS.520.4235S, 2024MNRAS.531.2168S}. Uniquely, the vAPP coronagraph preserves an image of the target star for use as a photometric reference, while simultaneously producing a coronagraphic dark hole in which high-contrast companions can be detected \citep[e.g.][]{2012SPIE.8450E..0MS, 2014OExpr..2230287O, 10.1117/12.2056096, 2021ApOpt..60D..52D, 2021MNRAS.506.3224S, 2023A&A...674A.115L, 2023AJ....165..216L, 2025A&A...698A..52M}. Such coronagraphs are installed on numerous ground-based imagers, with the photometric reference provided either as the main stellar PSF itself (with its coronagraphic dark hole) or as a separate, fainter, stellar PSF positioned at an offset, depending on the phase design of the specific vAPP \citep[see][]{2021ApOpt..60D..52D}. The vAPP produces the same PSF pattern for all sources in the field, including any companions. By further combining the vAPP with an integral field spectrograph (IFS), the light from both the host star and the companion are dispersed into spectra, which can then be extracted through aperture photometry and recombined to obtain a white-light time series for each object. This step helps to minimise the impact of any wavelength-specific flat-fielding errors, improving the precision of the light curves compared to broad-band photometric observations. A differential light curve for the companion can then be produced by dividing the companion flux by that of the host star, thereby eliminating trends arising from systematics shared by both objects, leaving behind only non-shared variations. However, while this includes the intrinsic variability of the companion, any remaining systematics not shared by the star and companion also remain. These can be further corrected to some extent, where their sources are known; in a pilot study of this technique, \citet{2023MNRAS.520.4235S} used a parametric linear regression approach to fit and remove residual trends from sources such as airmass, achieving a 3.7\% precision per 18-minute bin in their differential light curve of substellar companion HD~1160~B. Similar studies of transiting exoplanet transmission spectroscopy and secondary eclipses often correct for non-shared systematics using more comprehensive polynomial models or Gaussian processes \citep[e.g.][]{2011A&A...528A..49D, 2012MNRAS.419.2683G, 2018AJ....156...42D, 2023AJ....165..169D, 2019A&A...631A.169T, 2022MNRAS.510.3236P, 2022MNRAS.515.5018P}. Yet, understanding the sources and magnitudes of the systematics that impact light curves obtained through ground-based differential spectrophotometry is key to accurately estimating the precision achieved with this method, and for devising new approaches to mitigate these systematics and hence reach greater precision in the future.

In this paper, we assess the extent to which telluric and instrumental systematics contaminate the differential light curves obtained with the technique of vAPP-enabled ground-based differential spectrophotometry. We do this by injecting artificial companions with and without variability to real data to test the shapes of the recovered light curves, and by producing simulated data to explore the impact of specific systematics. We use the instantaneous PSF of the host star as the template for the artificial companion injections to capture time-varying systematics that would impact a real companion, with the caveat that this approach does not account for systematics arising from differences in colour between the star and companion. In some cases, this may lead to optimistic conclusions about recoverability and precision. Nonetheless, artificial companion injection is an effective way to assess the extent to which unknown systematics limit the precision that we achieve with this technique. Meanwhile, simulated data allows us to measure the strength of some of the systematics that we are aware of, such as those caused by uncorrected aberrations described by Zernike modes. In Section~\ref{p4_ch5_planet_injection}, we describe the methods used to inject the artificial companions, process the data, and produce differential white-light curves for each companion. The simulated data is described in Section~\ref{p4_ch5_simulations}. In Section~\ref{p4_ch5_results}, we test how well the injected variability signals are recovered. We discuss these results and their implications for the light curve precision in Section~\ref{p4_ch5_discussion}, and lastly summarise the conclusions of this work in Section~\ref{p4_ch5_conclusions}.
\section{Artificial companion injection}\label{p4_ch5_planet_injection}
We can assess the level of variability that can be recovered in this type of data, and whether it can be recovered consistently at different locations in the data, by injecting artificial companions with simulated variability signals into real observational data. Furthermore, injecting companions with no variability (i.e. a flat signal) allows us to test the extent to which the differential light curves are affected by systematics. Such tests are not possible using real companions, as their level of variability is usually not known a priori, and itself can change over time \citep[e.g.][]{2022AJ....164..239Z, 2024ApJ...965..182F, 2024ApJ...970...62P}. In this section, we inject artificial companions with and without variability signals into an observational dataset, reduce the data, and produce differential light curves for these companions following the standard method used for real companions by \citet[][]{2023MNRAS.520.4235S, 2024MNRAS.531.2168S}. We then compare the recovered variability signal to that which was originally injected.

\subsection{Ground-based differential spectrophotometry method}\label{p4_ch5_method_steps}

In this subsection, we briefly summarise the ground-based differential spectrophotometry method for measuring the variability of high-contrast companions using a vAPP coronagraph, as initially presented by \citet[][]{2023MNRAS.520.4235S}. Additional detail related to the specific dataset used here is described in the following subsections where relevant.

\begin{figure*}
\centering
  \textcolor{white}{\frame{\includegraphics[scale=0.58]{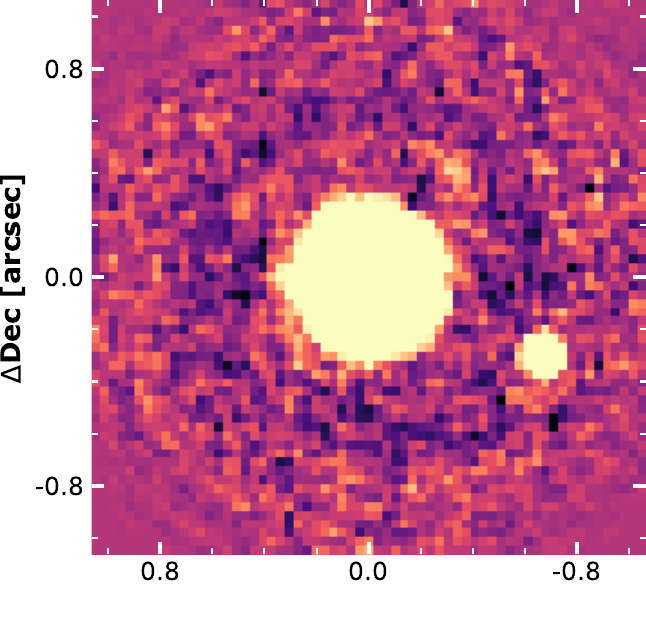}}}%
  \textcolor{white}{\frame{\includegraphics[scale=0.58]{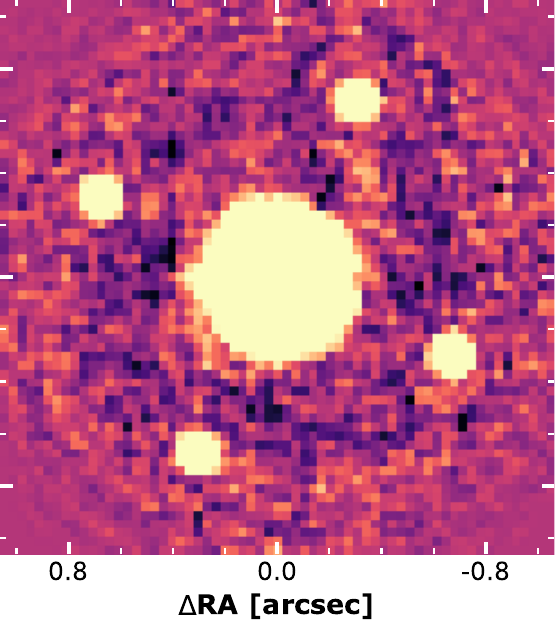}}}%
  \textcolor{white}{\frame{\includegraphics[scale=0.58]{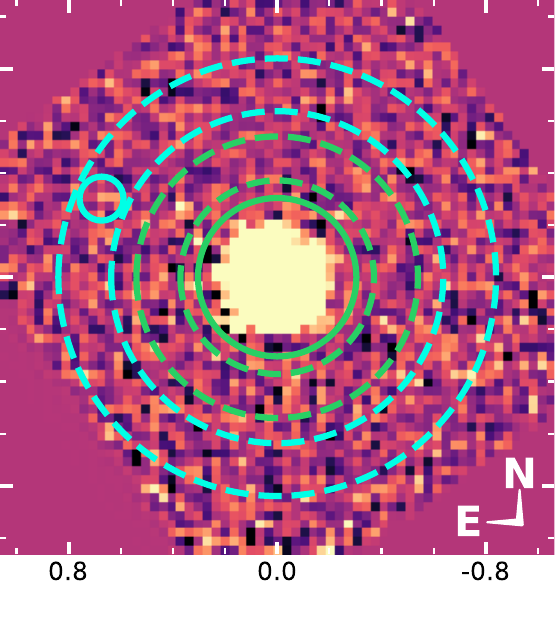}}}%
\caption{The left-hand and centre panels are examples of the final processed LBT/ALES+dgvAPP360 images produced when the data were median-combined in both time and wavelength. Left: the case where no artificial companions were injected to the data, so only the bright host star HD~1160~A and its bona fide companion HD~1160~B is visible. Centre: similar to the left-hand panel, but three artificial companions have been injected at 90\textdegree{} intervals in position angle from HD~1160~B. All three artificial companions were injected with contrasts of $2.88$~$\times$~10$^{-3}$ (6.35~mag) relative to the host star. This image is a composite; for the purposes of the analysis, only one companion was injected at a time. Right: A single frame of data highlighting examples of the apertures (solid lines) and annuli (dashed lines) used to extract photometry and background measurements for the host star (in green) and artificial companions (in blue). The left-hand and centre panels use the same arbitrary logarithmic colour scale while the right-hand panel uses a different one, and all three panels are aligned to north, where north is up and east is to the left.}
\label{fig:p4_ch5_real_injected_comparison}
\end{figure*}

\begin{enumerate}
  \item[(i)] Firstly, a star with a high-contrast companion is observed using a vAPP coronagraph in conjunction with an IFS. The vAPP enables the companion and its host star to be observed simultaneously, while the IFS allows (spectro)photometry to be obtained across many wavelength channels. Nodding is used to facilitate background subtraction.
  \item[(ii)] Background subtraction is then performed using data from the alternative nod position. The spectra are then extracted into 3D image cubes of spatial position and wavelength.
  \item[(iii)] Standard data reduction steps are applied, such as bad pixel correction and flat-field correction. Frames with AO loop issues are removed, as are problematic wavelength channels such as those affected by absorption by a glue layer in vAPP coronagraphs \citep[$\sim$3.25-3.5~\textmu m, ][]{2017ApJ...834..175O, 2021ApOpt..60D..52D}. The frames are spatially and rotationally aligned.
  \item[(iv)] Photometric measurements are taken for both the star and the companion for all frames in both wavelength and time. \citet[][]{2023MNRAS.520.4235S} used apertures centred on the star and companion and corresponding annuli for background measurements (e.g. Figure~\ref{fig:p4_ch5_real_injected_comparison}, right-hand panel). The white-light flux measurements for each object are then obtained by taking the median combination in wavelength, reducing the impact of wavelength-specific systematics and improving the signal-to-noise ratios (S/N) of the star and companion.
  \item[(v)] Next, a differential white-light curve is obtained by dividing the white-light photometry of the companion by that of the star. This step aims to eliminate systematic trends shared by both objects from the light curve of the companion.
  \item[(vi)] Additional detrending is then performed to mitigate residual systematics that are not shared by the star and companion. The approach used by \citet[][]{2023MNRAS.520.4235S} was a multiple linear regression approach using airmass, air temperature, wind speed, wind direction, and the pixel positions of both the star and companion. The resulting light curve is the detrended, differential, white-light curve of the companion.
  \item[(vii)] This companion light curve is then binned to the required cadence and precision.
\end{enumerate}

\subsection{Ground-based differential spectrophotometry dataset}\label{p4_ch5_datasets}
The vAPP-enabled differential spectrophotometric monitoring dataset used here for our artificial planet injection and recovery tests is that presented by
\citet{2023MNRAS.520.4235S}, who conducted a variability study of substellar companion HD~1160~B. This dataset was obtained on the night of 2020 September 25 (03:27:31 - 11:16:14 UT) with the left-side aperture of the 2~x~8.4-m Large Binocular Telescope (LBT) in Arizona, using the double-grating 360\textdegree{} vector Apodizing Phase Plate \citep[dgvAPP360;][]{2017SPIE10400E..0UD, 2020PASP..132d5002D, 2021ApOpt..60D..52D} coronagraph. The Arizona Lenslets for Exoplanet Spectroscopy (ALES) IFS was used with an L-band prism to spectrally disperse the light from the target over a 2.8-4.2~\textmu m wavelength range with an R$\sim$40 spectral resolution \citep[ALES;][]{2015SPIE.9605E..1DS, 2018SPIE10702E..0CS, 2018SPIE10702E..3LH, 2018SPIE10702E..3FS, 2022SPIE12184E..42S}. ALES works alongside the LBT Mid-InfraRed Camera (LMIRcam) as part of the LBT Interferometer (LBTI), providing a 2.2$\arcsec$x 2.2$\arcsec$ field of view with a $\sim$35~mas~spaxel$^{-1}$ plate scale \citep[][]{2010SPIE.7735E..3HS, 2012SPIE.8446E..4FL, 2016SPIE.9907E..04H, 2020SPIE11446E..07E, 2024SPIE13095E..06I}. \citet{2023MNRAS.520.4235S} obtained $\sim$3.32 hours of integration time on the HD~1160 system over $\sim$7.81 hours using an on/off nodding pattern, with 109.7\textdegree{} of field rotation and stable weather conditions. Importantly, companion host star HD~1160~A has been shown to be non-variable to the 0.03\% level using data from the Transiting Exoplanet Survey Satellite (TESS) mission, making it suitable for use as a simultaneous photometric reference \citep{2023MNRAS.520.4235S}. To avoid any issues arising from micro-spectra overlap (i.e. flux contamination from the neighbouring spaxels in the dispersion direction that can impact data close to the start and end of the wavelength range) and the dgvAPP360 glue absorption feature at $\sim$3.25-3.5~\textmu m, we chose to use a subset of this data set covering a wavelength range of 3.59-3.99~\textmu m for our analysis \citep{2017ApJ...834..175O, 2021ApOpt..60D..52D}.

To enable artificial companions to be injected to the data easily, the raw ALES micro-spectra grids were first converted into 3D image cubes of spatial position and wavelength according to the procedure described by \citet{2023MNRAS.520.4235S}; once the sky background had been subtracted using the data obtained in the off-source nod position, the micro-spectra were extracted using weighted optimal extraction \citep{1986PASP...98..609H, 2018SPIE10702E..2QB, 2019AJ....157..244B, 2020AJ....160..262S}. Wavelength calibration of the micro-spectra was carried out using four fiducial spots provided by narrow-band filters located upstream of ALES which were fitted using a second-order polynomial, allowing pixel position to be mapped to wavelength \citep{2018SPIE10702E..3FS, 2022SPIE12184E..42S}. The final image cube consisted of 30 wavelength channels, with 2200 frames per channel.

\subsection{Injecting artificial companions}\label{p4_ch5_inject_process}
Artificial companion injection is widely used in high-contrast imaging studies as a method for obtaining photometric measurements of bona fide companions \citep[e.g.][]{2010Sci...329...57L, 2010SPIE.7736E..1JM, 2011A&A...528L..15B, 2016ApJ...820...40A}. This is generally done using an unsaturated PSF of the host star, obtained separately, which acts as the artificial companion. The brightness of the real companion is then measured by subtracting the artificial companion at its location in the images, while iteratively scaling the artificial companion's brightness until the residuals at this location are minimised. Here, we apply the concept of artificial planet injection to insert additional companions into the images, but instead use the instantaneous PSF of host star HD~1160~A provided by the dgvAPP360 in each frame as the template for the artificial companion in that frame. This is usually not possible for high-contrast imaging data as the host star is often blocked by a focal-plane coronagraph in such observations \citep[e.g.][]{2012SPIE.8442E..04M, 2018SPIE10698E..2SR}. However, this novel frame-dependent approach is advantageous because the template PSFs will reflect frame-to-frame changes, caused by time-varying systematics, in the shapes and sizes of the PSFs of real companions and their host stars. This has not been possible in previous studies and is a unique advantage of the vAPP coronagraph.

We produced the artificial companion template PSF for each frame by first duplicating the frame, then dividing it by a flat frame produced by combining frames from the off-source nod position in the same wavelength channel. We then cropped the template to a 12-pixel radius, and shifted it to the image coordinates where we wished to inject an artificial companion using a spline interpolation approach. Both the rotation of the field and drifts in the position of the star on the detector were taken into account in calculating these coordinates, such that the companion was injected at the desired separation and position angle relative to the star. Next, we set all pixels less than 1\% of the peak flux to zero, and scaled the flux to the required star-companion contrast (see below). Where we wanted to simulate a variability signal, we did this by further multiplying the template by a sinusoidal function with the corresponding amplitude, period, and phase. Finally, we multiplied the template by the flat frame again and added it to the original data frame to inject the companion.

We injected six companions, three with no variability and three with simulated sinusoidal variability. Only one companion was injected per iteration to maximise the size of the region available for the subtraction of residual background flux (see Section~\ref{p4_ch5_data_reduction}). We did not attempt to remove the real companion HD~1160~B from the data, but ensured that the artificial companions were physically separated from it by choosing position angles at 90\textdegree{}, 180\textdegree{}, and 270\textdegree{} offset from that of HD~1160~B. We used the physical separation of HD~1160~B ($\sim$0.78$\arcsec$) for the separation of the artificial companions, as this placed them centrally in the coronagraphic dark hole of the dgvAPP360. We also used the flux of HD~1160~B as a baseline flux for many of the injected companions, assuming an L'-band contrast of $\Delta L^{\prime}=6.35$ mag (or $2.88$~$\times$~10$^{-3}$) \citep{2012ApJ...750...53N}. \citet{2023MNRAS.520.4235S} found sinusoidal-like variations in their light curve of HD~1160~B and fitted them with a 8.8\% semi-amplitude sinusoid with a period of 3.239~h, phase shift of 0.228, and y-offset of 0.993. To enable a comparison to their results, we simulated this variability signal for the time-varying artificial companions.

\subsection{Data processing and extracting spectrophotometry}\label{p4_ch5_data_reduction}
Once an artificial companion had been injected to the data, we followed the standard steps for processing data of this type and extracting photometry of the targets, as described by \citet[][]{2023MNRAS.520.4235S}.

Firstly, we corrected for errors in the response of the detector by dividing each frame by the flat frame previously used in the preparation of the artificial companion templates. We then masked the host star HD~1160~A, the companion HD~1160~B, and the artificial companion, before fitting and removing a third-order polynomial from each image column and then repeating this process for each row. This was done to correct for systematic discontinuities that exist in ALES data, arising from the overlap of the micro-spectra with different LMIRcam detector channels \citep{2022AJ....163..217D}. We then shifted the frames to align the star to the centre of each frame using a spline interpolation approach, and derotated them to account for the field rotation and align the data to north. Examples of the final images are shown in the left-hand and centre panels of Figure~\ref{fig:p4_ch5_real_injected_comparison}, median combined in time and wavelength to highlight the companions by increasing their S/N. The left-hand panel shows the final image with no artificial companions; the real companion, HD~1160~B, can be clearly seen. This is the same as the left-hand panel of Figure~3 in \citet[][]{2023MNRAS.520.4235S}. The centre panel then additionally contains three artificial companions with the same contrast as HD~1160~B, but located at positions offset from it by intervals of 90\textdegree{} in position angle. This image is a composite; in practice, only one artificial companion was injected into the data at a time, but we show multiple artificial companions per frame here to demonstrate the relative locations at which they were injected.

Next, we extracted aperture photometry for the host star and each artificial companion in each frame in both wavelength and time. We used the same aperture radii as \citet[][]{2023MNRAS.520.4235S}, which were 9 pixels (3.1~$\lambda$/D) and 2.5 pixels (0.9~$\lambda$/D) for the star and artificial companions, respectively. We also subtracted any residual background flux in these apertures using the annuli to estimate the background at their locations. For the star, we did this using an annulus centred on the star with an inner radius of 11 pixels and an outer radius of 16 pixels. For the companions, we also used an annulus centred on the star, but with a width of 6 pixels at the radial separation of the companion. Both the artificial companion and HD~1160~B were masked for this process, so that they did not contaminate our estimate of the background. An example of these apertures and annuli for an artificial companion 180\textdegree{} offset from HD~1160~B can be seen in the right-hand panel of Figure~\ref{fig:p4_ch5_real_injected_comparison}, superimposed on a single frame of data.

\subsection{Companion light curves}\label{p4_ch5_lcs}
Once we had extracted photometric measurements for the host star and each artificial companion in each wavelength channel, we applied the steps of \citet[][]{2023MNRAS.520.4235S} (see Section~\ref{p4_ch5_method_steps}) to create detrended differential white-light curves for each companion. We first took the median combination of the photometric measurements over the 3.59-3.99~\textmu m range, producing single white-light flux measurements for each object at each time. For each companion, we then divided these white-light flux measurements by the white-light flux measurements of the host star. This step has the effect of removing any systematic trends shared by the time series of both objects from the flux of the companion, leaving behind a differential light curve containing non-shared variations only. This includes both the simulated variability signal of the injected companion and any residual systematic trends. Such systematics can arise from the effects of Earth's atmosphere, as well as from the instrumentation and data reduction process, and many of these systematics will be the result of differences in the properties of the star and companion \citep[e.g.][]{2005AN....326..134B, 2006MNRAS.373..231P}. However, we note that while the artificial companions that we inject here do differ from the host star in brightness and position in the data, they do not reflect the difference in colour that would exist for a real companion because their template PSFs were constructed using the PSF of the star. In this regard, the artificial companions are not perfectly reflective of true companions and thus these residual systematics may differ slightly.

\begin{figure*}
	\includegraphics[width=\textwidth]{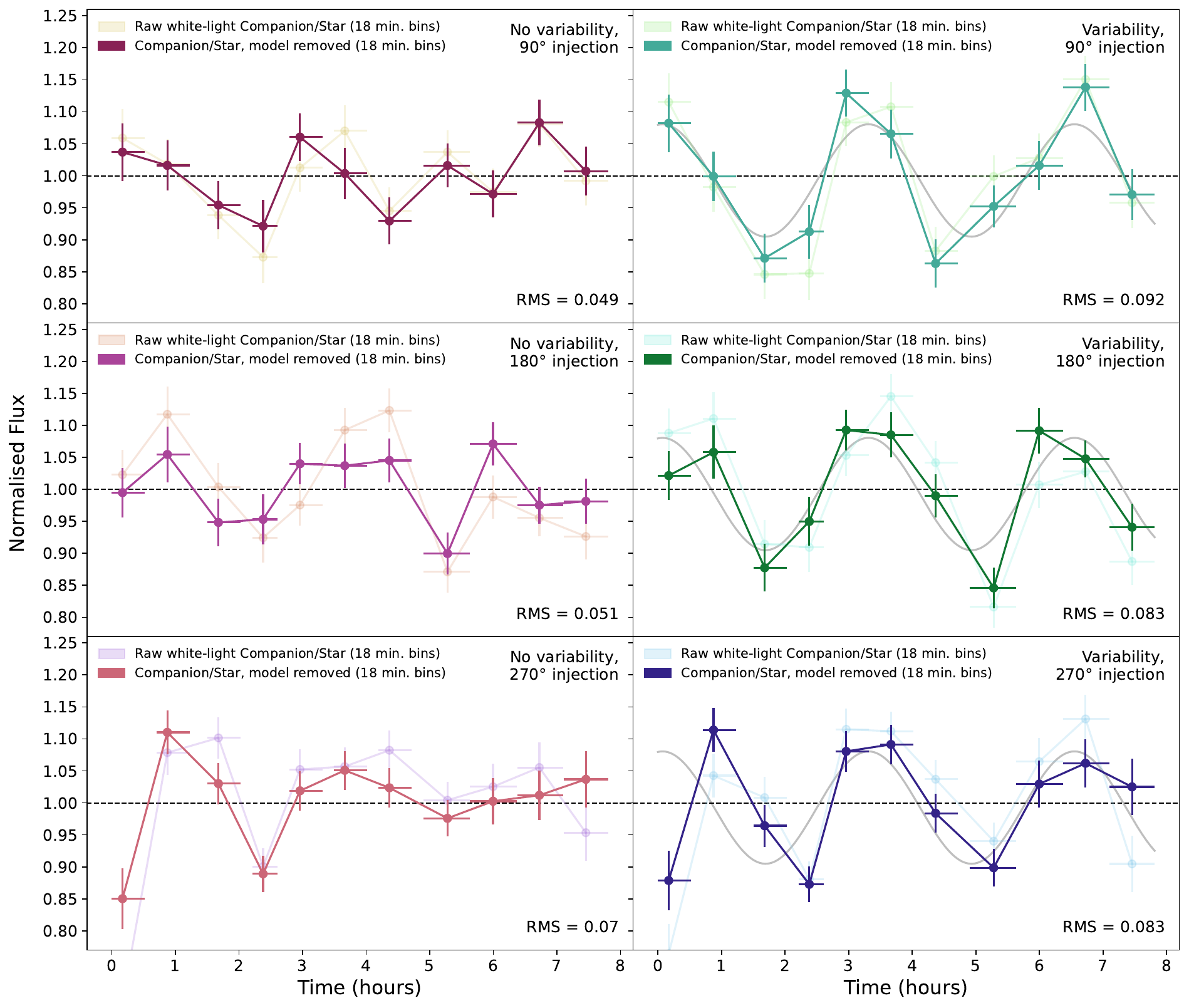}
    \caption{The raw differential white-light curves for each of the injected artificial companions are shown in lighter colours in each panel, binned to 18 minutes of integration time per bin. The detrended differential white-light curves, after division by the multiple linear regression model to remove the modelled systematic trends, are then overplotted in darker colours. The left-hand panels show the light curves for the artificial companions injected without variability, whereas the right-hand panels are those injected with a simulated sinusoidal variability signal, which is overplotted in grey for comparison. The root mean square (RMS) shown are those of the detrended light curves.}
    \label{fig:p4_ch5_injected_lcs}
\end{figure*}

Nonetheless, we proceeded to detrend the differential white-light curves of each artificial companion further using a multiple linear regression approach with the same decorrelation parameters used by \citet[][]{2023MNRAS.520.4235S}, thereby partially mitigating any residual systematics from these known sources. These decorrelation parameters were airmass, air temperature, wind speed, wind direction, and the x,y-positions of both the star and companion in the original data cubes. We produced a linear regression model for each artificial companion using these parameters, and then divided the model out of the light curve of the corresponding companion. The final, detrended, differential white-light curves for each artificial companion are shown in Figure~\ref{fig:p4_ch5_injected_lcs}, binned to 18~minutes of integration time per bin. The left-hand panels are those that were injected with no variability signal, and the right-hand panels are those that were injected with the variability signal found by \citet[][]{2023MNRAS.520.4235S} for HD~1160~B. The raw differential white-light curves, prior to detrending and again in 18~minute bins, are also shown overplotted in lighter colours for comparison. The flux error bars are the median absolute deviation (MAD) $\times$ 1.48 of the data points in each bin divided by $\sqrt{N-1}$, where $N$ is the number of frames in each bin. $N$ is the same for all bins; the differences in the time error bars arise from gaps in the data produced by the on/off nodding pattern used for the observations. We analyse the detrended differential white-light curves in Section~\ref{p4_ch5_results}.

\section{Simulations of LBT Data with HCIPY}\label{p4_ch5_simulations} \subsection{Simulating the LBT/ALES+dgvAPP360 system}\label{p4_ch5_lbt_dataset_sims}
\begin{figure}
    \centering
    \includegraphics[width = \linewidth]{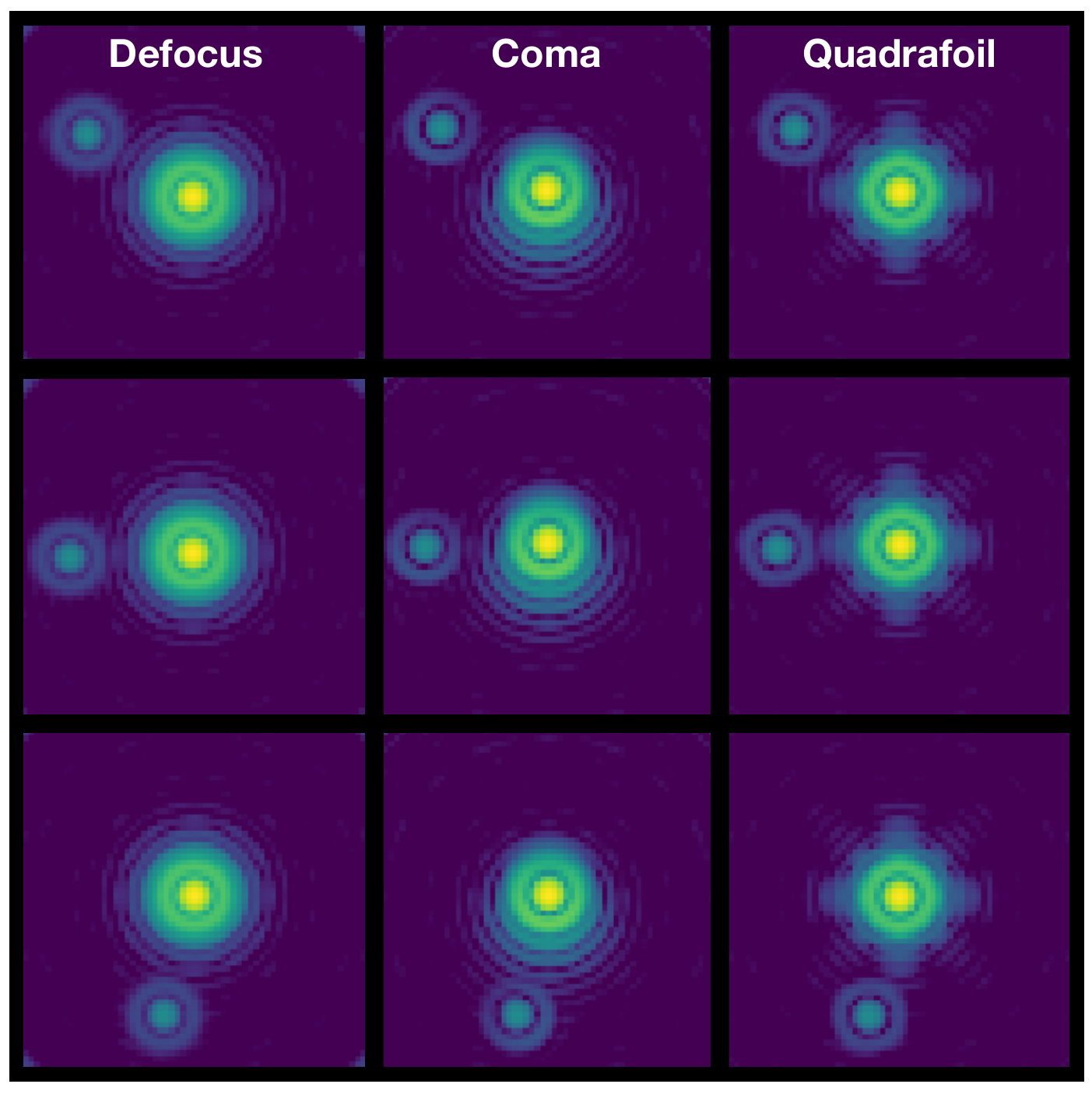}
    \caption{Simulated PSFs of a star and companion for a different Zernike mode are shown in each column at three different observing times. The data are in pupil-stabilised mode, so the aberrations remain static while the companion rotates over time. The symmetry in the Zernike mode and the location of the companion together determine the measured companion flux in an aperture.}
    \label{fig:p4_ch5_zernike_psfs}
\end{figure}
\begin{figure*}
    \centering
    \includegraphics[width = \linewidth] {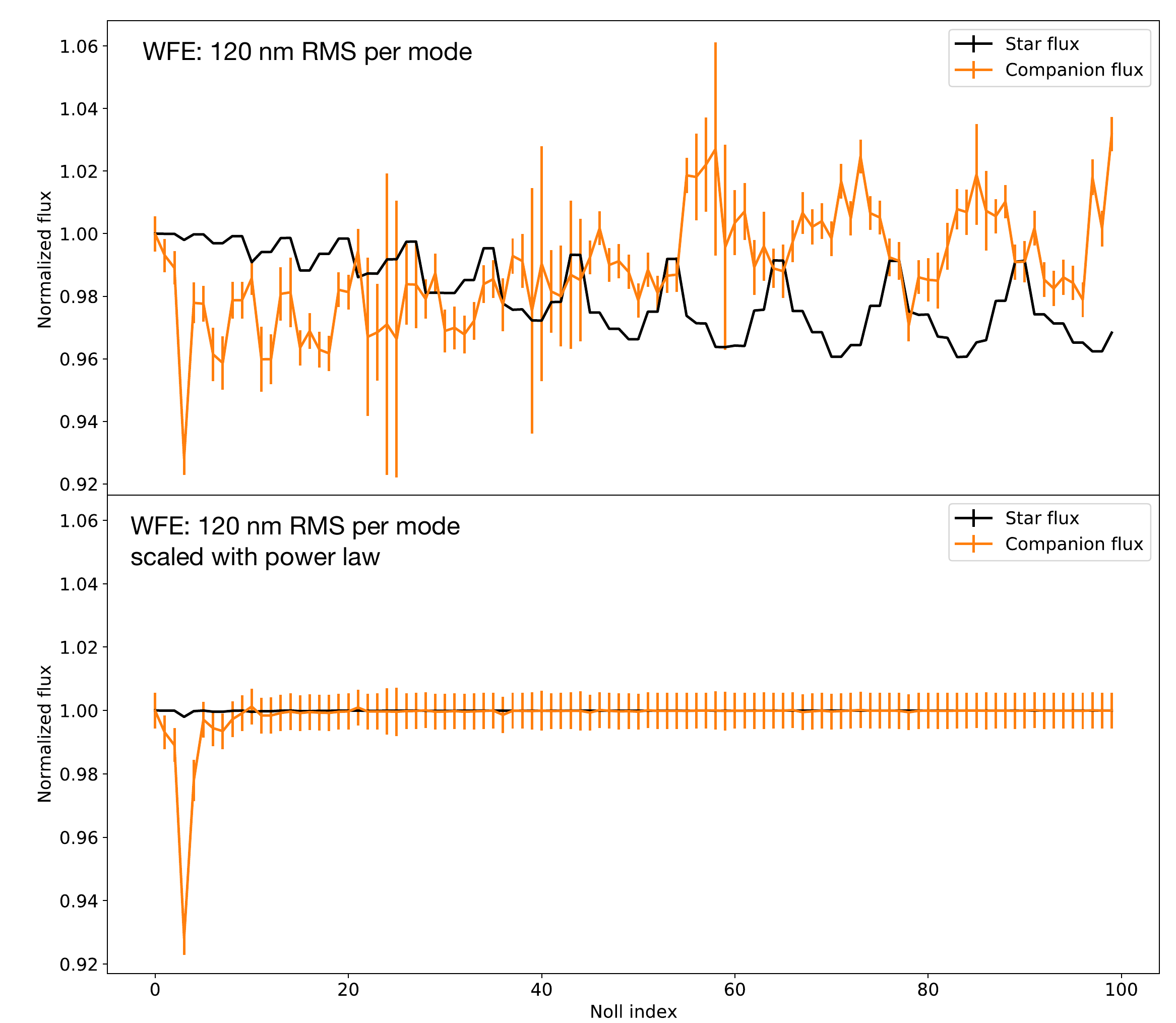}
    \caption{Simulated normalized flux of the star (black) and the companion (orange) for the first 100 Zernike modes, given by their Noll indices. The errors bars indicate the maximum and minimum retrieved companion flux over the observing sequence. The error bars for the star are too small to be visible. Top panel: all Zernike modes have the same 120~nm RMS wavefront error. Bottom panel: The power in the modes is scaled by a power law with a slope of -1.5, similar to expected NCPAs. In this more realistic scenario, we find that only the lowest-order aberrations are likely to significantly impact the average flux of the companion over the observing sequence.}
    \label{fig:p4_ch5_var_zernike_flux}
\end{figure*}
\begin{figure*}
    \centering
    \includegraphics[width = \linewidth] {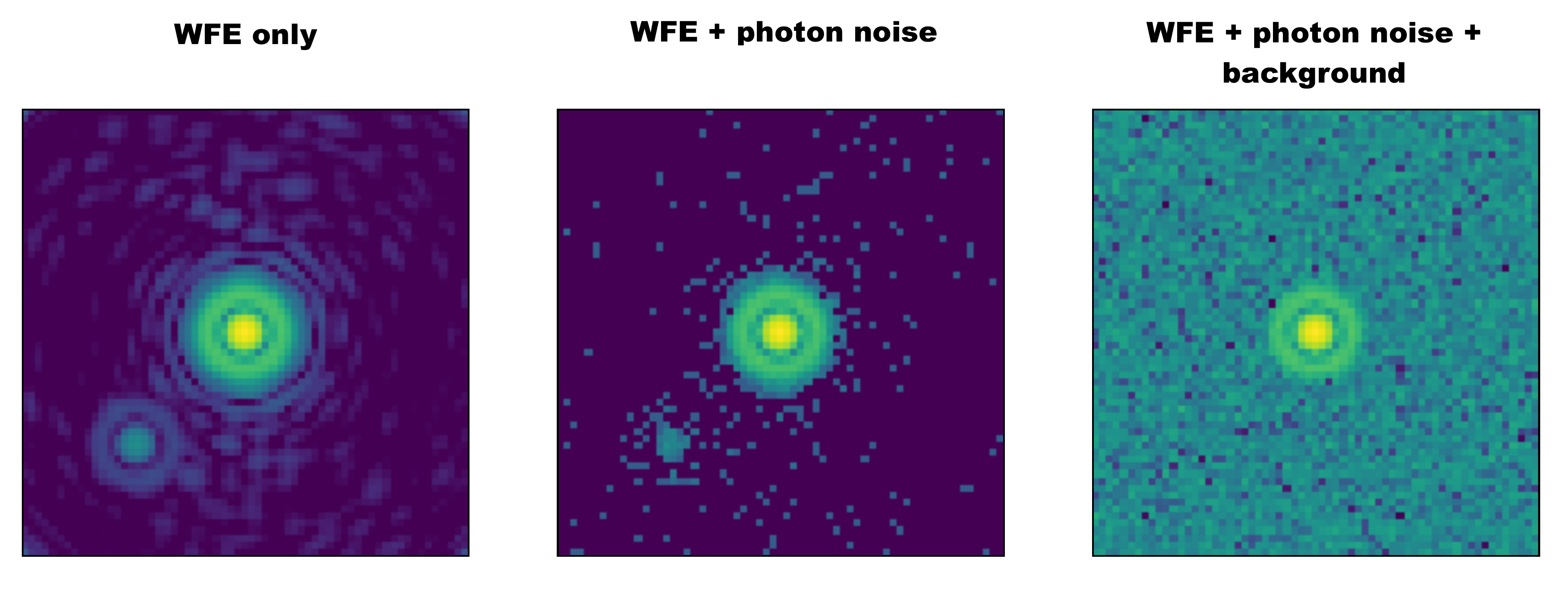}
    \caption{Simulated point-spread functions. The left-hand panel shows the PSF for the residual wavefront error after adaptive optics correction. The centre panel shows the same PSF simulated with photon noise, assuming a photon flux of 40.000 photons for a single image. This photon number is empirically matched to the counts in a single frame of the HD~1160 data. The right panel shows the same PSF as the left panel with photon noise and noise from the thermal background. This background noise is implemented as read noise and the amount is also empirically matched with the statistics in a single frame of the HD~1160 data.}
    \label{fig:p4_ch5_PSF_noise_comparison}
\end{figure*}
\begin{figure*}
    \centering
    \includegraphics[width = \linewidth] {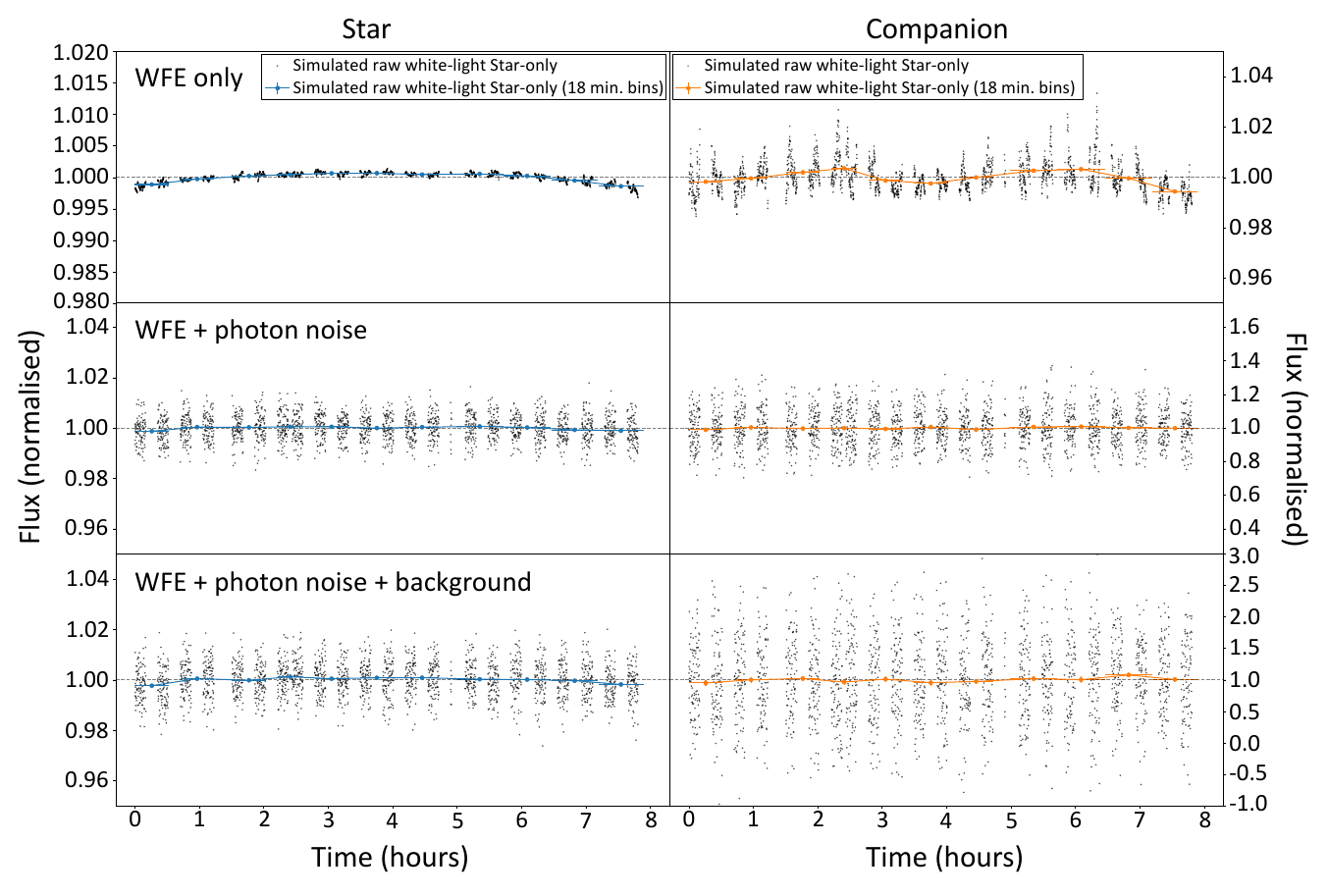}
    \caption{Simulated aperture photometry measurements for the star (left panels) and the companion (right panels) for the three aberration and noise scenarios.  }
    \label{fig:p4_ch5_variability_vs_noise}
\end{figure*}
\begin{figure*}
    \centering
    \includegraphics[width = \linewidth] {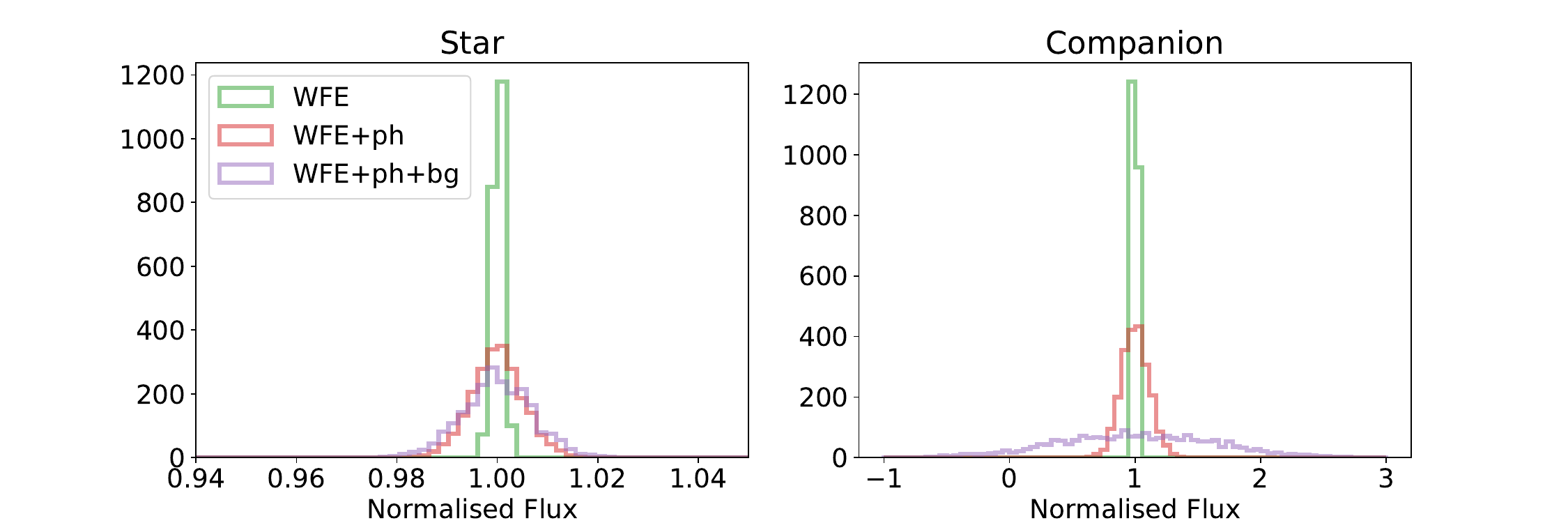}
    \caption{Histograms of the simulated photometric measurements for the star in the left panel and the companion in the right panel. The three colours indicate the three aberration and noise scenarios; residual wavefront error after adaptive optics correction, photon noise, and noise due to the thermal background.}
    \label{fig:p4_ch5_var_hist}
\end{figure*}
\begin{figure}
    \centering
    \includegraphics[width = \linewidth] {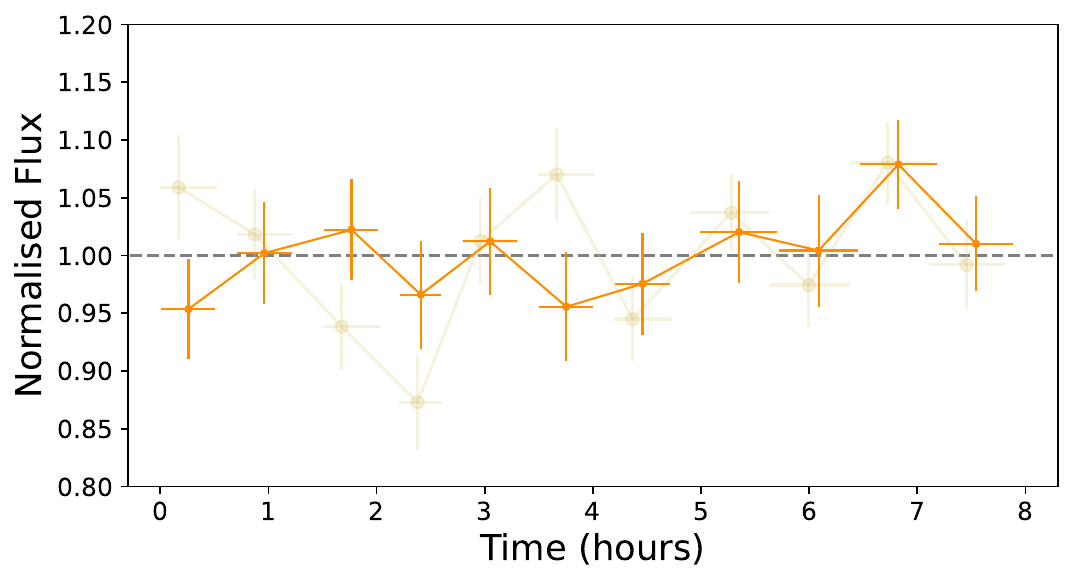}
    \caption{The orange line is the raw differential light curve for the (non-variable) simulated companion in the scenario that includes the residual wavefront error after adaptive optics correction, the photon noise, and the noise due to the thermal background. For comparison, we also show the raw differential light curve of the 90\textdegree{} artificial companion injected to the real dataset with no variability in Section~\ref{p4_ch5_planet_injection} (lighter line, reproduced from the upper left-hand panel of Figure~\ref{fig:p4_ch5_injected_lcs}).}
    \label{fig:p4_ch5_combined_sim_planet_variability}
\end{figure}
Artificial planet injection and recovery allows us to characterise the overall effect of systematics present in the data on differential light curves produced through the process described in Section~\ref{p4_ch5_planet_injection}. In this section, we take an additional step to understand the individual contributions of known sources of systematic errors: non-common path aberrations (NCPAs) and residual wavefront errors generated by the correction of atmospheric turbulence using adaptive optics (AO), i.e. AO residuals. Both effects can generate varying speckles at the location of a companion, influencing the flux measured with aperture photometry.

We used the Python package HCIPy \citep{2018SPIE10703E..42P} to produce simulated LBT/ALES+dgvAPP360 data including NCPA and AO residuals, allowing us to test their impact on variability measurements obtained using vAPP-enabled differential spectrophotometric monitoring. HCIPy is capable of generating both static wavefront errors and dynamic turbulence phase screens, simulating adaptive optics systems, propagating aberrations through the coronagraph to the focal plane, and simulating realistic camera images from the resulting PSF. HCIPy also correctly models coupled effects between NCPAs and AO residuals, and propagates them through to the simulated data. We simulated the LBT/ALES+dgvAPP360 system in HCIPy by propagating an unpolarized wavefront through the dgvAPP360 optic to the focal plane. This input wavefront had an amplitude given by the LBT pupil without secondary support and with no phase aberrations. We downsampled the dgvAPP360 design by a factor of 4.3 using matrix Fourier transforms to improve the simulation speed while minimizing the impact on the performance of the coronagraph. The focal plane sampling was chosen to closely match that of a single wavelength channel of LBT/ALES data after it has been extracted into a 3D cube of 63 x 63 pixel images. We directly compared our simulations to the background-subtracted images of the HD~1160 system described in Section~\ref{p4_ch5_datasets}, allowing us to match this sampling to real data to the sub-pixel level. These simulations are monochromatic at a wavelength of 3.75~\textmu m.

We then simulated a companion using the same steps, except that the input wavefront was given an additional tip and tilt phase ramp to place the source off-axis at the desired companion location. The location of the companion was matched with the photometric mask (i.e. aperture) used by \citet{2023MNRAS.520.4235S} to extract the flux of HD~1160~B. We also scaled the companion flux level to match that of HD~1160~B, assuming an L'-band contrast of $\Delta L'$ = 6.35 mag (or $2.88$~$\times$~10$^{-3}$) \citep{2012ApJ...750...53N}. We did not provide the simulated companion with a variability signal. As the simulated and real data were closely matched in this way, we were able to perform aperture photometry for the simulated star and companion using the same focal plane aperture masks used for the original analysis (see right-hand panel of Figure~\ref{fig:p4_ch5_injected_lcs}). However, we first injected the desired aberrations (e.g. NCPA and AO-residual wavefront aberrations) to the data to simulate their effect on the photometric measurements. These are described in the following subsections. As the dgvAPP360 coronagraph is a pupil-plane coronagraph, incoming flux is suppressed at all wavelengths equally well and the PSFs of all sources in the field are modified in the same way. Observations with the dgvAPP360 coronagraph are therefore unaffected by the chromatic and throughput effects that impact observations obtained with focal-plane coronagraphs \citep[e.g.][]{2025ARA&A..63..179K}. Its position in the pupil plane also makes it highly stable and inherently insensitive to tip/tilt instabilities arising from telescope vibrations \citep{2017ApJ...834..175O, 2022AJ....163..217D}.

\subsection{Impact of low-order aberrations}\label{p4_ch5_sim_zernike_modes}
NCPAs are aberrations generated by the optical system after the beam splitting of incoming light into the two paths that lead to the AO wavefront sensor and to the detector, respectively. These aberrations also vary in time due to effects such as atmospheric turbulence, thermal drifts, and vibrations in the instrumentation, with timescales ranging from a few seconds to several hours \citep[e.g.][]{2007JOSAA..24.2334S, 2014SPIE.9148E..5HN, 2019A&A...629A..11V, 2022A&A...660A.140V, 2022A&A...659A.170S}. In principle, slowly varying NCPAs could induce a false variability signal in differential light curves obtained using differential spectrophotometry, if they impact the extracted photometry of the star and the companion differently. To the first order, the dgvAPP360 coronagraph is insensitive to these aberrations as the impact on the Strehl ratio is the same for both the star and the companion. However, changes in the shapes and the sizes of the companion and star PSFs over time can impact the ratio of their fluxes, particularly if different aperture sizes are used for each object. The companion may also move over stellar speckles caused by NCPAs as the field rotates, contaminating its flux. Mitigating NCPAs is challenging, as they are introduced after the incoming light is split by the beam splitter and therefore cannot be corrected even by the most powerful AO systems. Furthermore, the properties of NCPAs cannot be inferred from observational data itself as the photon noise from the thermal background is too high. Here, we used HCIPy to investigate which aberrations have the largest effect on companion and stellar photometry and whether or not a realistic distribution of NCPAs can have a significant impact on variability measurements obtained through vAPP-enabled differential spectrophotometry.

We added simulated NCPAs to our simulated data using the first 100 Zernike modes, a series of polynomials that describe wavefront aberrations in optical systems \citep{1934Phy.....1..689Z, 1976JOSA...66..207N, 2022JOpt...24l3001N}. We added one mode per iteration, allowing us to measure their individual impact on the companion flux over the observing sequence. The Zernike modes were scaled to 120~nm RMS in the pupil. We varied the companion location by rotating it according to the 109.7\textdegree{} of field rotation of the real data set described in Section~\ref{p4_ch5_datasets}. However, the aberrations remained static with respect to the pupil as the observations were pupil-stabilised. We show the impact of three low-order aberrations (defocus, coma and quadrafoil) on the star and companion PSFs at three different observing times in Figure~\ref{fig:p4_ch5_zernike_psfs}. These images highlight how even a static aberration can affect the observed flux of a companion over an observing sequence as it moves over the spatially-varying structure of the stellar PSF. Symmetric modes will inherently induce less systematic variability, while asymmetric modes will have a greater impact. For example, variability induced by the quadrafoil aberration will have a higher frequency than that of the coma aberration. All variability induced by static modes is a direct function of the angular rotation rate, and is thus observatory dependent for a given object (and vice versa).

We then derotated the data by the rotation angles and extracted stellar and companion photometry for each of the 100 simulated Zernike modes, where the modes were all scaled to the same 120~nm RMS wavefront error. The time-averaged normalised fluxes of each object are shown as a function of Zernike mode \citep[represented by its Noll index,][]{1976JOSA...66..207N} in the top panel of Figure~\ref{fig:p4_ch5_var_zernike_flux}. The error bars indicate the minimum and maximum measured fluxes (i.e. the peak-to-peak variability amplitude arising from the aberrations) over the observing sequence covering the 109.7\textdegree{} of field rotation. The error bars for the star are too small to be visible. For the stellar flux, we find a decrease in flux and larger differences per mode for higher Noll indices. This is the direct result of scaling by RMS wavefront error, as higher-order modes will have a larger peak-to-valley error for the same RMS wavefront error. Interestingly, the measured companion fluxes do not match the same pattern as the stellar fluxes. Dividing the companion flux by the stellar flux therefore does not improve the photometric stability.

In this scenario, where the wavefront error is 120~nm RMS for a single mode, the offset of the companion flux from a normalised flux of one is on the order of a few percent for most Zernike modes. The speckles generated by a single Zernike mode dominate the measured companion flux for this planet-star contrast $\Delta L'$. The level of variability in the measured flux of the companion, arising from the changing rotation angle, also changes significantly between modes. We find that most modes induce $\sim$1\% variability over the observing sequence, most likely due to rotation and derotation interpolation effects which could also be present in real data of this type. However, the change in companion flux is much higher for some modes, up to $\sim$10\%.

At first glance, this paints a worrisome picture for the determination of companion variability in the presence of static NCPAs. However, the outcome is different if we consider a more realistic system. While the total residual wavefront error for high-contrast imaging systems can be on the order of 120 nm RMS \citep[e.g.][]{2014SPIE.9148E..5QH, 2016SPIE.9909E..52M, 2019A&A...621A...4R}, the wavefront error per mode is generally not as strong as assumed in the simulations above. The aberrations described by Zernike modes are expected to follow an inverse power law in aberration strength, and therefore quickly reduce in amplitude with increasing Noll index \citep[e.g.][]{2007JOSAA..24.2334S, 2018SPIE10703E..5ML}. We therefore repeated our simulations testing the impact of individual Zernike modes, this time applying a power law with a slope of -1.5 (a conservative estimate) as a function of radial frequency. The resulting time-averaged normalised fluxes for the star and the companion in this scenario are shown in the bottom panel of Figure~\ref{fig:p4_ch5_var_zernike_flux}. These more realistic simulations indicate that only the lowest-order aberrations are likely to significantly impact measurements of companion flux.

\subsection{Realistic simulations of vAPP-enabled differential spectrophotometry data}\label{p4_ch5_sim_ao_residuals}
In addition to our analysis of NCPAs, we also attempted to produce a more realistic simulation of the observational dataset targeting the HD~1160 system described in Section~\ref{p4_ch5_datasets}. With this goal, we used HCIPy to generate several noise factors including wavefront aberrations arising from atmospheric turbulence. Uncorrected wavefront aberrations can produce a varying field of residual stellar speckles that can impact companion variability measurements in much the same way as NCPAs \citep[e.g.][]{2007ApJ...654..633H, 2012A&A...541A.136M, 2013A&A...554A..41M, 2021PASP..133j4504M}.

First, we used the wind speed, wind direction, and airmass measurements obtained by \citet{2023MNRAS.520.4235S} for the HD~1160 data set to generate representative turbulence phase screens. We set the seeing to 1.1$\arcsec$, the coherence time 15 ms, and we scale the Fried parameter with the airmass. We simulated the adaptive optics system using the HCIPy adaptive optics layer, with 500 Zernike modes and a lag of two frames. The seeing and the AO loop speed were chosen such that a Strehl ratio of around 85\% was achieved in H-band and 98\% at 3.7~\textmu m, similar to the performance reported by \cite{2014SPIE.9148E..0LS}. We generated 100 random realizations simulating Earth's atmosphere and run the AO-system for 22 frames in 0.4 seconds, thereby producing 2200 frames, the same number as the HD~1160 data set. This allowed us to match each frame with a frame from the HD~1160 data and move the companion according to the rotation angle of that frame. We did not add NCPAs for this simulation as the impact of these aberrations on variability measurements was found to be small in the previous section, and we now wished to test the larger effects that dominate the scatter in our photometry.

Next, we made the frames more reflective of real data by adding photon noise and background noise using the NoisyDetector module of HCIPy. We calculated the photon noise for each frame using a total power of 40,000 photons, which we matched empirically to the count levels of the HD~1160 data. The background noise comes from the photon noise of the thermal background; as the HD~1160 frames were background-subtracted using the data obtained in the off-source nod position before the multi-wavelength image cubes were extracted, it is difficult to estimate the actual background levels for our simulated wavelength channel. We therefore chose to include the photon noise of the background through the read noise option of the NoisyDetector module, empirically matching the noise levels to the HD~1160 data as 12 counts. Example frames from these simulations showing the PSFs of the star and the companion are shown in Figure~\ref{fig:p4_ch5_PSF_noise_comparison}. The left-hand panel shows the PSFs when we only include the residual wavefront error due to uncorrected atmospheric turbulence in the simulation. This turbulence generates speckles in the wind direction, which add up to form a faintly visible wind-driven halo \citep[e.g.][]{2018A&A...620L..10C, 2020A&A...638A..98C, 2019JATIS...5d9003M}. The excellent performance (i.e. high Strehl ratio) of the simulated adaptive optics system in the L-band means that residual speckles are minimal and are less bright than the companion. The centre panel is the same as the left-hand panel, but with photon noise added to the simulation. Now, only the core of the companion is visible due to the low number of counts. Finally, in the right-hand panel, we also add background noise. The companion is no longer visible in a single frame as the background dominates its signal at its location. This is consistent with the real HD~1160 observations, for which HD~1160~B cannot be seen in a single frame. The background also significantly contributes to the measured stellar flux.

We then extracted photometry for the star and the companion in each frame for each of the three scenarios shown in Figure \ref{fig:p4_ch5_PSF_noise_comparison}, again using the same photometric masks after derotating the frame. These normalised fluxes are shown as a function of time in Figure~\ref{fig:p4_ch5_variability_vs_noise}. The left-hand panels are the fluxes of the star and the right-hand panels are those of the companion. We find that the star varies by less than 1\% in the residual wavefront error case. This variation shows a clear trend and can be attributed to the reduced AO performance for the larger airmasses at the start and end of the observing sequence. In this case, the companion flux shows higher amplitude trends with a more complex shape than a simple airmass correlation. However, its variability is nonetheless only on the order of 1\%. When photon noise is added (second row panels), the stellar flux shows a significantly increased scatter. Moreover, the scatter in the companion flux is between $\pm 30\%$ and binning is required to recover precise photometry. Finally, in the case where all three noise factors are included (bottom row panels), the scatter in the stellar flux is not much greater than before. However, the scatter in the measured companion flux has increased dramatically and sometimes even negative flux values are measured. The scatter in the companion flux that is generated in the simulation has a similar magnitude to the scatter in the raw companion flux in the HD~1160 data set as measured by \citet{2023MNRAS.520.4235S}, whereas the scatter of the simulated stellar flux is less than that of the raw stellar flux in the HD~1160 data. However, this simulation has several key caveats and is not a perfect reflection of real data. There are several important effects that we did not include in the simulation, such as the reduction of atmospheric transmission with increasing airmass.

We further compared the normalised flux distributions of the 2200 simulated frames in all three scenarios, allowing us to explore the respective contributions of each noise source to the measured photometry. These are shown in Figure~\ref{fig:p4_ch5_var_hist}, where the left-hand and right-hand panels show the histograms of the stellar and companion fluxes, respectively. If we consider the stellar flux, we see that the stellar flux measurements are dominated by the photon noise of the star itself. The background noise has a far more dramatic impact for the companion and dominates the recovered photometry, as we might intuitively expect from the PSFs in Figure~\ref{fig:p4_ch5_PSF_noise_comparison}. Both photon noise and the background noise are random, and so binning the frames will reduce the scatter with the square root of the number of frames per bin (see Section~\ref{p4_ch5_results}). This behaviour of the noise was also demonstrated for the HD~1160 data set by \citet{2023MNRAS.520.4235S}, and can be seen following the white noise trend in their Figure~13.

To allow a comparison to the differential light curves of the injected companions shown in Figure~\ref{fig:p4_ch5_injected_lcs}, we divided the simulated companion flux for the scenario including all three noise sources by the simulated stellar flux, thereby producing a raw differential light curve. We then binned this light curve to the same binning used for the light curves of the injected companions in Section~\ref{p4_ch5_lcs} (i.e. 11 bins of 200 frames per bin). The obtained raw differential light curve is shown in Figure~\ref{fig:p4_ch5_combined_sim_planet_variability}, and has a scatter and error bars that closely match those of the non-varying artificial companions shown in Figure \ref{fig:p4_ch5_injected_lcs}. This suggests that the dominant effects in the real data have been accurately accounted for in our simulations.
\begin{figure*}
	\includegraphics[width=\textwidth]{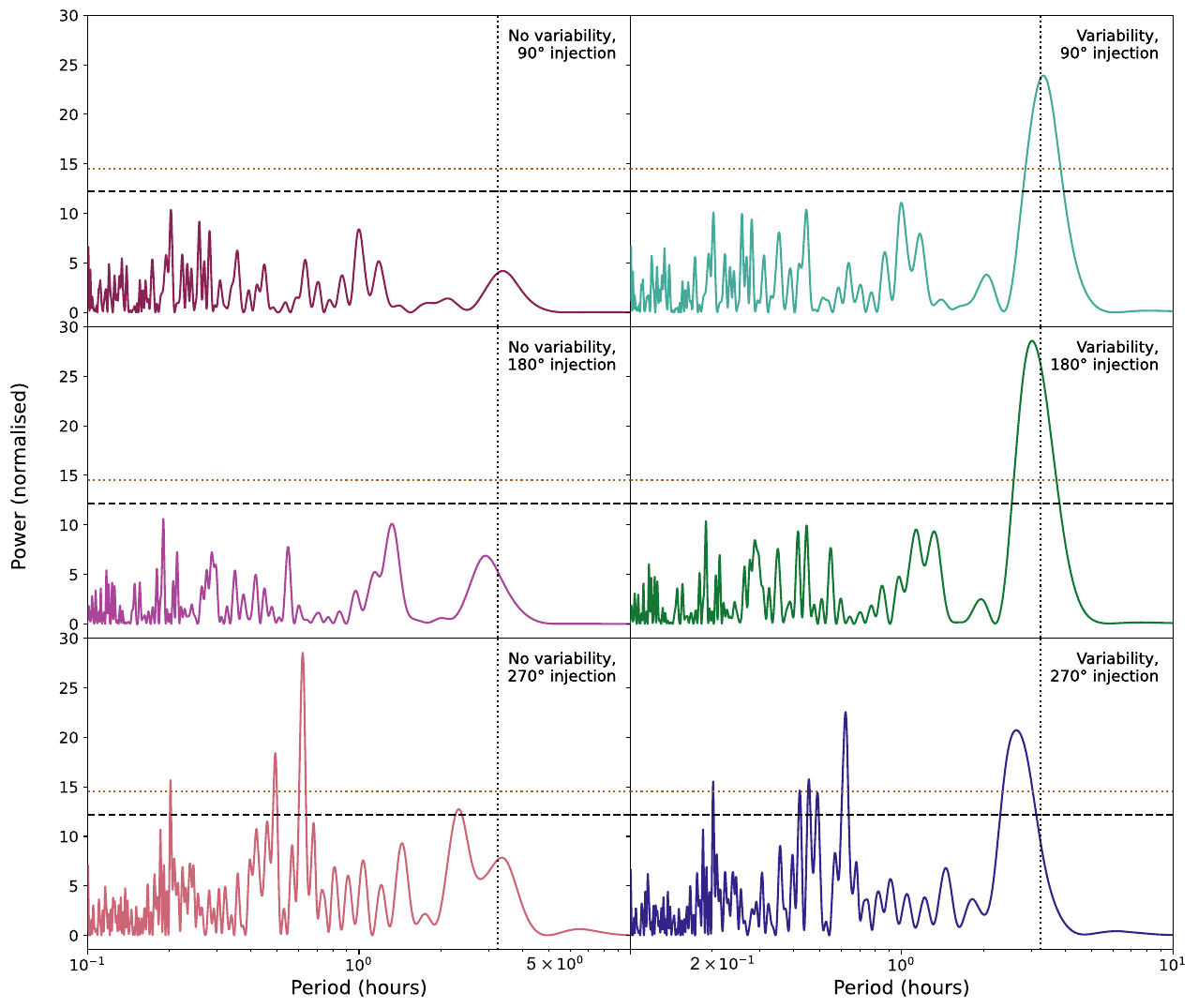}
    \caption{The Lomb-Scargle periodograms for the differential white-light curves of each artificial companion. The left-hand panels are the periodograms for the companions injected without any variability, while the right-hand panels are those for the companions injected at the same coordinates but with a sinusoidal variability signal. The vertical dotted lines indicate the $\sim$3.24~h period of the injected variability signal. The horizontal black dashed lines and brown dotted lines show the power thresholds corresponding to false-alarm probabilities of 0.1 (10\%) and 0.01 (1\%), respectively.}
    \label{fig:p4_ch5_periodograms}
\end{figure*}
\section{Artificial companion variability analysis}\label{p4_ch5_results}
In this section, we search for periodic signals in the detrended differential white-light curves of the artificial companions. This allows us to test not only whether we can recover the variability properties that were used for the time-varying artificial companions, but also whether residual systematics induce false periodic trends in the light curves of the non-varying artificial companions. We then use the light curves of the non-varying companions to further assess the limiting precision of this data set.

We produced Lomb-Scargle periodograms for each artificial companion using their unbinned detrended differential white-light curves \citep{1976Ap&SS..39..447L, 1982ApJ...263..835S}. Each periodogram was normalised by the variance of the datapoints in the corresponding light curve, following the implementation of \citet{1986ApJ...302..757H}. The periodograms for the artificial companions that were injected without and with variability are shown in the left-hand and right-hand columns of Figure~\ref{fig:p4_ch5_periodograms}, respectively, horizontal black dashed lines and brown dotted lines representing the 10\% and 1\% false-alarm power thresholds, respectively. The powers, periods, and false alarm probabilities of the strongest peaks in the periodograms for each of the companions injected with a variability signal are given in Table~\ref{table:p4_ch5_var_properties}. We find no peaks above 1$\sigma$ for the artificial companions injected without any variability at 90\textdegree{} and 180\textdegree{} offsets from HD~1160~B, as would be expected for a flat light curve. For the artificial companions injected with variability at the same positions, we find $\sim$5-6$\sigma$ peaks at approximately the injected period. However, the periodograms for the companions injected at a 270\textdegree{} offset are more surprising. The periodogram of the non-varying companion at this location shows a strong peak at a 0.619~hour period, with several other peaks above the 1\% false-alarm power threshold. The strongest peak in the 270\textdegree{} time-varying companion periodogram is also at this period, albeit with a lesser power. This may indicate that the light curves of the injected companions at this location are contaminated with one or more short-period periodic systematics. The second strongest peak in the periodogram of the time-varying companion does lie close to the injected period, although it has a shorter period of 2.638~hours. 

\begin{figure*}
	\includegraphics[width=\textwidth]{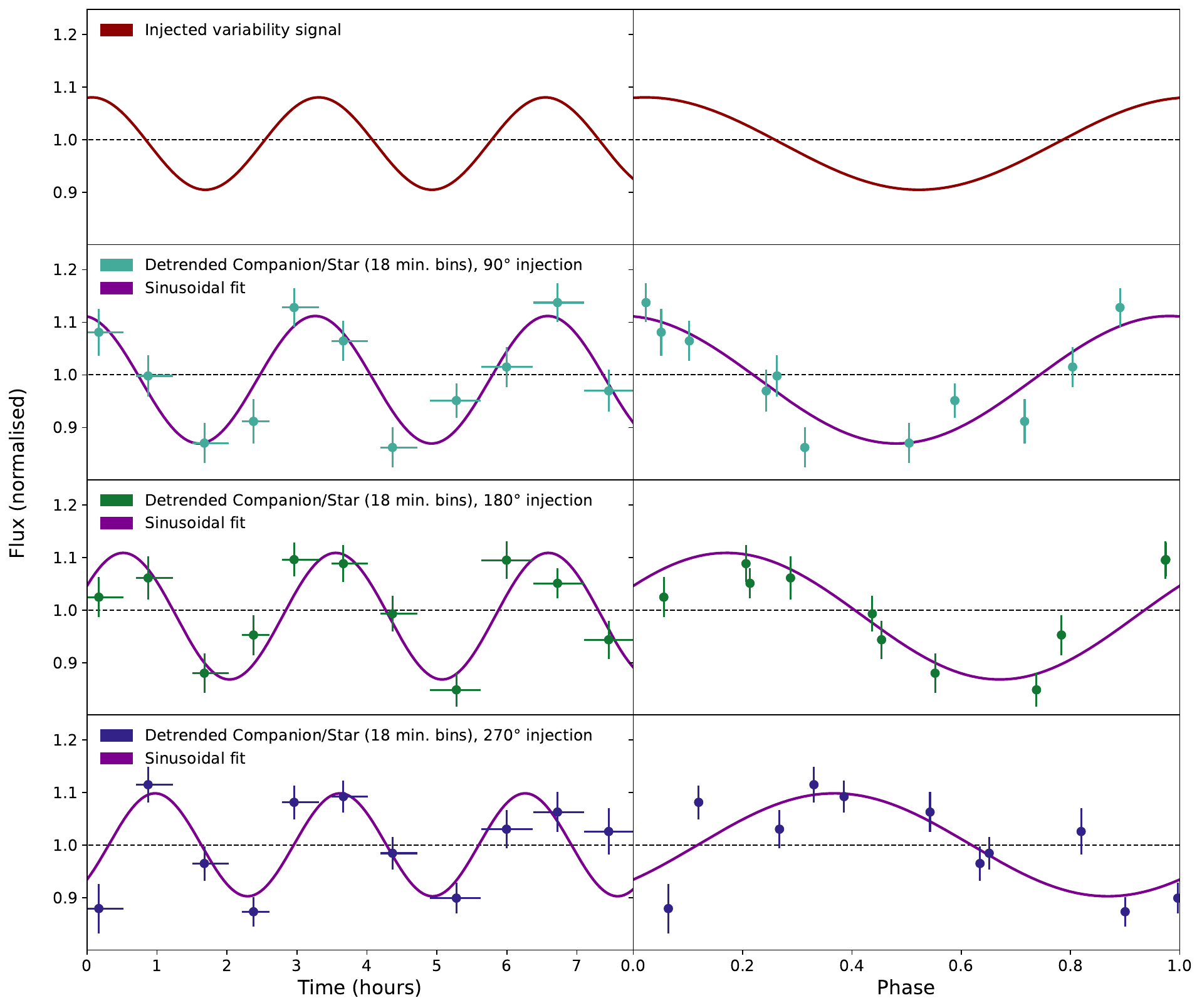}
    \caption{The top row shows the sinusoidal variability signal that was given to the artificial companions injected with variability, as a function of time in the left-hand panel and phase-folded to its 3.24~hour period on the right. The left-hand panels of the following three rows show the detrended differential white-light curves of the three artificial companions that were injected with this variability, reproduced from the right-hand column of Figure~\ref{fig:p4_ch5_injected_lcs}, and the purple lines show the best-fitting sinusoids to these light curves. These light curves and sinusoids are then phase-folded to their respective periods in the right-hand panels.}
    \label{fig:p4_ch5_sine_fits}
\end{figure*}

We further fitted sinusoids to the detrended differential white-light curves of the time-varying artificial companions so that we could directly compare the amplitude and phase of their variability to that of the injected signal. We did this using a non-linear least squares approach with the 3.34~hour and 3.03~hour periods obtained from the periodogram peaks as the initial guesses for the fits to the 90\textdegree{} and 180\textdegree{} light curves, respectively. For the 270\textdegree{} light curve, we used the 2.64~hour period of the second strongest peak in its periodogram as the initial guess, assuming that this peak does arise from the variability signal that we injected. The properties of these sinusoidal fits are given in Table~\ref{table:p4_ch5_var_properties}, and the fits themselves are shown overplotted in purple on the corresponding light curves in the left-hand column of Figure~\ref{fig:p4_ch5_sine_fits}. The top panel of this figure shows the original injected variability signal, for comparison. The panels in the right-hand column are the same as on the left, but phase-folded to the periods of the respective sinusoids. We can see that while the recovered sinusoids are broadly similar to the injected variability, their amplitudes are consistently slightly higher than what was injected. The phases of the recovered variability signals are also different, although these values appear to be correlated with the recovered period values such that the peaks and troughs of the injected and recovered sinusoids are roughly aligned. We discuss these results further in Section~\ref{p4_ch5_discuss_injections}.

\begin{figure}
	\includegraphics[width=\linewidth]{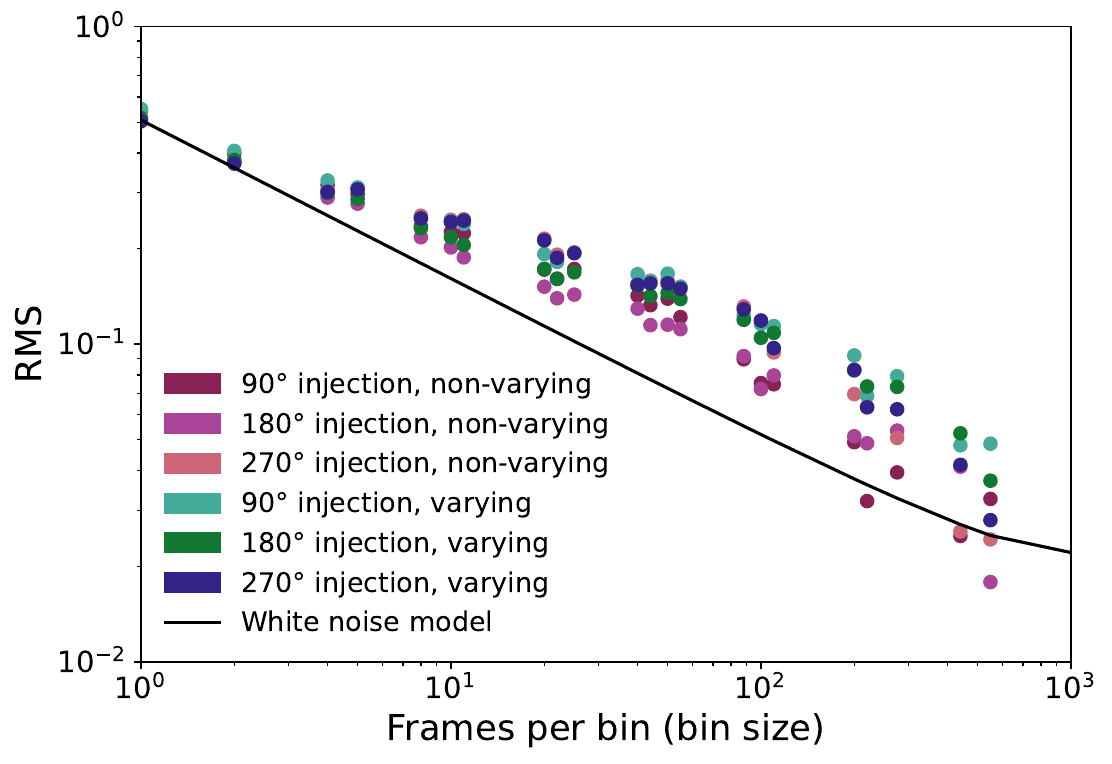}
    \caption{The RMS of the binned detrended differential white-light curves of the six injected artificial companions as a function of bin size. The theoretical white noise model is also shown.}
    \label{fig:p4_ch5_rms_plot}
\end{figure}

We also assessed the noise properties of each light curve using the method used by \citet{2011ApJ...733...36K} and \citet{2023MNRAS.520.4235S}. We first binned the unbinned detrended differential white-light curve of each artificial companion to a range of bin sizes, then normalised the data and subtracted a value of one to centre each light curve around zero. The RMS of each light curve was then measured for each bin size. These values are plotted in Figure~\ref{fig:p4_ch5_rms_plot}, with a black line showing the theoretical white noise model. We find that the RMS values of the light curves of the artificial companions injected with time variability sit higher than those without variability.

\begin{flushleft}
\begin{table}
\caption{The properties of the injected variability signal and the sinusoidal variability recovered from the detrended differential white-light curves of each of the time-varying artificial companions. The powers, periods, and false alarm probabilities of the strongest peaks in the periodogram for each of these companions are also given. The RMS values of the non-varying artificial companion light curves for a bin size of 200 frames per bin are shown in the bottom row.}
\begin{tabular}
{p{0.26\columnwidth}p{0.12\columnwidth}p{0.12\columnwidth}p{0.12\columnwidth}p{0.12\columnwidth}}
\hline
Variability&Injected&90\textdegree{}&180\textdegree{}&270\textdegree{}\\
property&variability&injection&injection&injection\\
\hline
Pgram. peak power&--&23.9&28.6&22.6\\
Pgram. peak period&--&3.337&3.027&0.622\\
Pgram. peak FAP&--&7.13e-07&5.99e-09&2.85e-06\\
\hline
Period&3.239&3.324&3.038&2.644\\
Semi-amplitude&0.088&0.121&0.120&0.098\\
Phase&0.228&0.269&0.080&-0.118\\
$y$-offset&0.993&0.991&0.988&1.001\\
\hline
\hline
RMS precision&--&0.0492&0.0513&0.0696\\
\hline

\end{tabular}
\label{table:p4_ch5_var_properties}
\end{table}
\end{flushleft}
\section{Discussion}\label{p4_ch5_discussion}
\subsection{Artificial companions}\label{p4_ch5_discuss_injections}
In Section~\ref{p4_ch5_results}, we produced Lomb-Scargle periodograms for the detrended differential light curves of each of the six injected artificial companions. We find that we successfully recover the expected variability signals for the companions injected at 90\textdegree{} and 180\textdegree{} offsets from the real companion HD~1160~B. We find no significant peaks for those injected with no variability signal (i.e. a flat line) at these locations, and detected strong peaks at the expected period for those that were injected with sinusoidal variability matching that of HD~1160~B \citep[as measured by ][]{2023MNRAS.520.4235S}. Although the injected variability was chosen to match the fitted sinusoid obtained for the variability of HD~1160~B, these periodogram peaks (5-6$\sigma$) are more significant than that measured for the periodic variability of HD~1160~B \citep[$\sim$2.5$\sigma$,][]{2023MNRAS.520.4235S, 2024MNRAS.531.2168S}. This could be because the intrinsic variability of HD~1160~B is complex and cannot be perfectly described by a singular sinusoid with a regular period. The variability signals of several other substellar companions in the literature have been attributed to multiple atmospheric features, and in many cases have been seen to evolve over time \citep[e.g.][]{2009ApJ...701.1534A, 2015ApJ...799..154M, 2016ApJ...825...90K, 2017Sci...357..683A, 2022AJ....164..239Z}. However, the strength of our recovery of the variability of our artificial companions is more likely a reflection of the caveats of our artificial companion injection approach. The PSF template that we used for the injected companions was produced using the instantaneous PSF of the star in a given frame. This is unique in direct imaging and highly advantageous, as it allows us to capture the frame-to-frame changes due to time-varying systematics that would impact a real companion. However, this also means that the companion has the same colour as the star, so the difference in colour between a star and a real companion are not taken into account. The artificial companions therefore do not suffer from additional systematics that would arise for two objects of different colours, such as their different response to changes in airmass, for which accurate detrending is key \citep[e.g.][]{2005AN....326..134B, 2022MNRAS.510.3236P}. Furthermore, the variability signal that we injected for the time-varying companions was the same in each of the 30 wavelength channels that were combined to produce their white-light fluxes. The variability of real substellar companions such as HD~1160~B is unlikely to be achromatic \citep[e.g.][]{2013ApJ...778L..10B, 2024MNRAS.532.2207B, 2016ApJ...826....8Y, 2019AJ....157..101M, 2020AJ....160...77Z, 2020ApJ...893L..30B, 2025MNRAS.539.3758C, 2025ApJ...981L..22M}.

While the 90\textdegree{} and 180\textdegree{} injected companions produced the expected results, those at an offset of 270\textdegree{} did not. Although a significant peak was detected close to the injected period for the time-varying companion at this location, a stronger peak was detected at a far shorter 0.619~hour period. Moreover, this peak and others above the 1\% false-alarm power threshold are also present in the periodogram of the companion injected with a flat signal. It is clear that the fluxes of the companions at this location are contaminated by periodic systematics. However, the cause of these systematics is not clear from the data. Although there are no unexpected features visible in the images, possible causes could include field-dependent aberrations such as detector bad pixels or stray light. The companion moves over different pixels as the field rotates, so low-level effects such as these could feasibly give rise to anomalous periodicity in its light curve. The detrending process used here attempts to fit and remove residual systematics caused by such effects by including the pixel positions of both the star and companion as decorrelation parameters, but may not have fully mitigated them in the case of the 270\textdegree{} injection. More sophisticated detrending approaches, such as those developed by the transmission spectroscopy community using Gaussian processes, may help to correct for such systematics further \cite[e.g.][]{2018AJ....156...42D, 2022MNRAS.510.3236P}. Another possible explanation for the anomalous peak at 0.619 hours is the irregular sampling of the data arising from the on/off nodding pattern. \citet{2024MNRAS.531.2168S} previously produced periodograms for the window functions of this data set, and found that this irregular sampling can lead to strong periodogram peaks at periods $<$1~h. As an additional test to investigate this anomalous behaviour further, we measured the `light curve' of the background at the location of the 270\textdegree{} injection by performing the same process without injecting a companion. We then searched for periodicity in this light curve and found that these features are not present when the light curve has been detrended by fitting and removing a linear regression model with new regression coefficients, but that they do appear if we search for periodicity when no detrending is applied. This rules out anomalies in the detrending process as the source of the peak at 0.619 hours, and further suggests that it is characteristic to this location in the data. This supports the theory that this is caused by a local systematic. However, it also shows that in this case the detrending procedure struggled to capture the effects of short period (<1~hr) systematics when an injected source was present.

If we consider the properties of the sinusoids that were fitted to the unbinned differential white-light curves (Table~\ref{table:p4_ch5_var_properties} and Figure~\ref{fig:p4_ch5_sine_fits}), we find that their periods and phases are different and that none are a perfect match for the injected sinusoid. This suggests that a single night of data is insufficient to accurately and reliably measure variability properties for variability of this period. The $\sim$7.81 hour duration of the data used here only covers $\sim$2.41 periods of the 3.239~h period of the injected variability, so it might be the case that a longer baseline covering more periods would achieve more consistent results. We also find that the recovered amplitudes are all greater than the injected amplitude. This may indicate that the approach of fitting simple sinusoids to light curves to measure variability tends to produce overestimated amplitudes.

\begin{figure}
	\includegraphics[width=\linewidth]{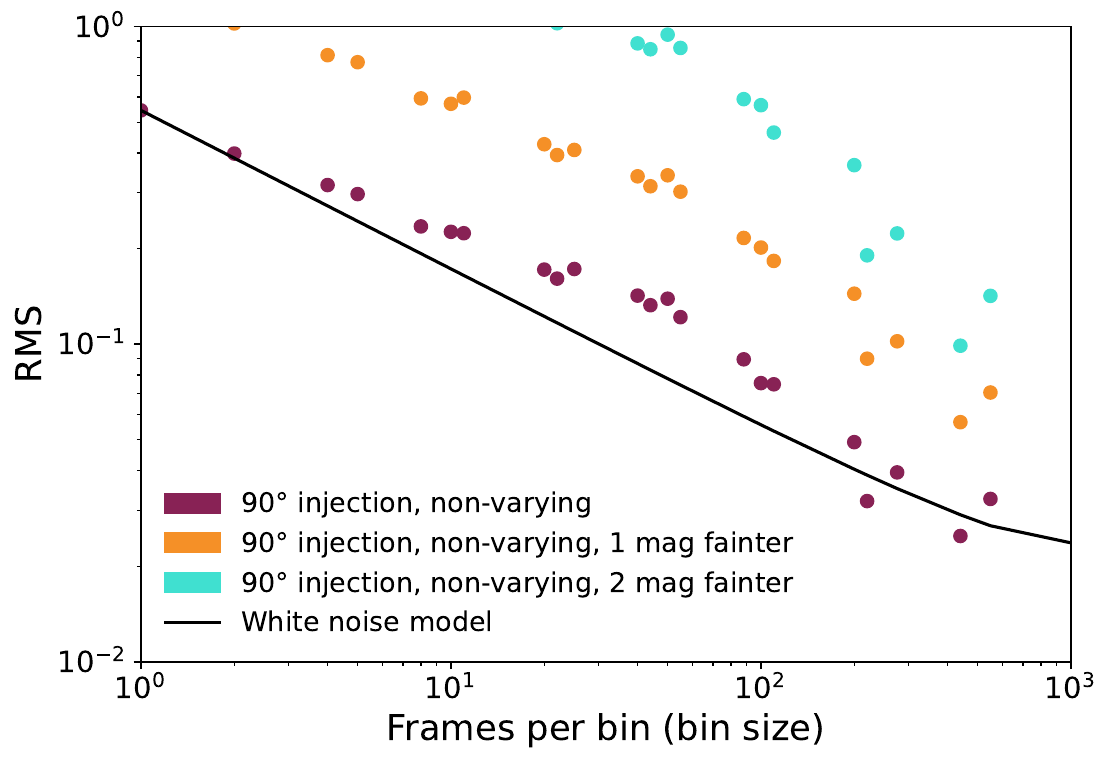}
    \caption{The RMS of the binned detrended differential white-light curves of the three non-varying artificial companions injected at the 90\textdegree{} injection position with different contrasts as a function of bin size. The theoretical white noise model for the brightest injection is also shown.}
    \label{fig:p4_ch5_fainter_rms_plot}
\end{figure}
\subsection{Light curve precision}\label{p4_ch5_discuss_precision}
We also assessed the noise properties of the detrended differential white-light curves of each injected companion by measuring their RMS for a range of bin sizes (Figure~\ref{fig:p4_ch5_rms_plot}). If we compare the RMS trend for the time-varying artificial companions to those injected without variability, we see that the RMS measurements of those with variability are generally higher, demonstrating that the level of scatter of the datapoints is distinct for companions with and without variability. We also see that these RMS trends do not plateau, matching the conclusion of \citet{2023MNRAS.520.4235S} that this data set has not yet reached a noise floor and that further increasing the bin size with additional data would lead to an even greater precision.

If we take the RMS values of the non-varying artificial companion light curves at the bin size used in Figure~\ref{fig:p4_ch5_injected_lcs} (200 frames per bin), we can produce a range for the limiting precision achieved at this bin size. These RMS measurements, given in the bottom row of Table~\ref{table:p4_ch5_var_properties}, are 0.0492, 0.0513, and 0.0696 for the 90\textdegree{}, 180\textdegree{}, and 270\textdegree{} companions, respectively. If we interpret these RMS values as the sensitivity limit (for the given bin size) in each case and consider their spread, we find a precision range of $\sim$4.9-7.0\%. We note that the RMS measurement for the 270\textdegree{} companion is higher than the other two, likely due to the greater impact of systematics in this light curve, and thus the upper extent of this range is higher. For comparison, the RMS values at this bin size for the corresponding varying companions are 0.0921, 0.827, and 0.0830. We can also compare these results to those found by \citet{2023MNRAS.520.4235S} for HD~1160~B. They measured $\sim$8.8\% semi-amplitude variability in their detrended differential white-light curve, with a precision of 3.7\% for a bin size of 200 frames. This precision was calculated by dividing the unbinned detrended differential white-light curve of HD~1160~B by the fitted sinusoid to remove the intrinsic variability of the companion, then taking the RMS value (0.037) of the resulting light curve when binned to 200 frames per bin. This 8.8\% variability is greater than the precision range estimated above and thus is likely to be astrophysical variability. The RMS values measured for the non-varying artificial companions here are broadly consistent with the precision measured by \citet{2023MNRAS.520.4235S} for HD~1160~B after the variability had been divided out, albeit slightly higher.

It is important to note that the sensitivity that can be achieved with the technique of vAPP-enabled ground-based differential spectrophotometry is dependent on the brightness and contrast of the target. Light curves measured for companions with a lower S/N will have an inherently higher scatter, making significant detections of low-level variability more challenging. However, it is also dependent on the bin size used, i.e. the achievable sensitivity is inversely proportional to light curve cadence, in the regime where the data are not systematic-limited. Thus, there is a trade-off between obtaining high-cadence light curves or binning them further to improve the sensitivity limit. The sensitivity limits discussed above are therefore specific to the case where the light curves for a companion such as HD~1160~B are binned to a specific cadence, whereas Figure~\ref{fig:p4_ch5_rms_plot} highlights this trade-off for a range of bin sizes. Although we do not derive a general sensitivity limit (or formula for evaluating this based on contrast, bin size, etc.) for this technique here using this single dataset alone, it may be possible to do so with additional datasets for a range of targets.

In an effort to conduct a preliminary inquiry into the sensitivity level that could potentially be achieved for fainter companions, we repeated the above analysis for two non-varying injected companions at fainter contrasts. These were injected at the 90\textdegree{} injection location, without variability, and with contrasts 1 magnitude and 2 magnitudes fainter than that of the previous injections, respectively. The same method described above was applied to these two companions to produce detrended differential white-light curves for each. We then measured the RMS of these light curves for a range of bin sizes to compare their RMS trends to that of the previous non-varying 90\textdegree{} injection. These trends are shown in Figure~\ref{fig:p4_ch5_fainter_rms_plot}, with that of the previous non-varying 90\textdegree{} injection reproduced from Figure~\ref{fig:p4_ch5_rms_plot}.

As one would expect, the RMS measurements for the light curves of the fainter companions are higher, indicating a lower sensitivity to variability. We measured the limiting precision of the fainter injected companions at a bin size of 200 frames per bin by taking the corresponding RMS measurements, as we did for the injections above. This yielded precisions of 14.4\% and 36.6\% for the 1~magnitude and 2~magnitudes fainter injections, respectively, compared to the 4.9\% obtained for the injection at full brightness. For comparison, \citet{2023MNRAS.520.4235S} detected variability in HD~1160~B with a semi-amplitude of 8.8\% with a precision of 3.7\% for the same cadence. This suggests that it would be challenging to detect variability with the same amplitude as HD~1160~B for companions even 1~magnitude fainter in contrast, unless a compromise is made on the cadence of the light curve to improve the precision. For example, if the bin size of the light curve for the 1~magnitude fainter companion is increased from 200 to 440, in this case an increase from 18~minutes to 39.6~minutes of integration time per bin, the precision improves to 5.7\%. Thus, a moderate decrease in cadence can lead to a significant improvement in precision. However, we note that unlike for the artificial companions injected with the same contrast as HD~1160~B, there is no real companion of similar contrast to the fainter injections to provide a robust baseline for comparison. We thus refrain from drawing definitive conclusions about the sensitivity that can be achieved for fainter companions using this single dataset alone.

\subsection{Non-common path aberrations}\label{p4_ch5_discuss_ncpas}
In Section~\ref{p4_ch5_simulations}, we produced a simulated differential spectrophotometry data set and added different aberrations to test their impact on the measured fluxes of the target PSFs.

We found that NCPAs have a significant impact on the measured fluxes of the star and the companion when we assume a wavefront error of 120~nm RMS per Zernike mode, but that this is significantly lessened when we scale each mode by a more realistic (yet conservative) power law with a slope of -1.5. We see from the bottom panel of Figure~\ref{fig:p4_ch5_var_zernike_flux} that in this latter case, high-order modes have minimal impact on the measured fluxes. However, some specific low-order modes can still lead to a reduction in the measured average companion flux over the observing sequence. This is because some modes (i.e. defocus, see Figure~\ref{fig:p4_ch5_zernike_psfs}) cause the flux of the target to spread out more than others, hence less flux is contained within an aperture of the same size. Despite this average reduction in flux for some modes, we note that the error bars for these modes are small ($\pm$<1\%) compared to some of those in the top panel. This suggests that for a given mode, the flux of the companion remains relatively stable over the observing sequence as it rotates through the field of view, and so the impact of these modes on companion variability measurements may be relatively minimal. In real differential spectrophotometry data, time variability arising from these modes is also likely partially mitigated by the detrending process applied to the differential white-light curves. In Section~\ref{p4_ch5_lcs}, we detrended the differential white-light curves of the artificial companions using a multiple linear regression approach including the positions of the companion and the star as decorrelation parameters. Thus, systematic trends associated with the movement of the companion over aberrations may be accounted for in the linear regression model.

For future observations with other instruments, systematics such as these could be further mitigated using wavefront sensing techniques that remove NCPAs, such as phase diversity \citep[e.g.][]{1982OptEn..21..829G, 1992JOSAA...9.1072P, 2017SPIE10400E..0UD, 2018SPIE10703E..1TM}. Several such approaches for removing NCPAs have already been demonstrated on sky, including focal-plane wavefront sensing techniques such as DrWHO \citep{2022A&A...659A.170S}, the phase diversity approach of Fast \& Furious wavefront sensing \citep[FnF, or sequential wavefront sensing, e.g.][]{2020A&A...639A..52B}, and coronagraphic correction with the ZELDA wavefront sensor \citep{2019A&A...629A..11V}. Predictive wavefront control algorithms that predict the evolution of atmospheric turbulence over short timescales will also help to reduce their impact \citep[e.g.][]{2017arXiv170700570G, 2019SPIE11117E..0WJ, 2020A&A...636A..81V, 2022JATIS...8b9006V, 2021JATIS...7b9001H, 2022SPIE12185E..82F}.

\subsection{Simulated noise sources}\label{p4_ch5_discuss_simulated_noise}
We also produced a separate simulation in which we attempted to simulate some of the key noise sources that affect real differential spectrophotometry data obtained with LBT+ALES/dgvAPP360; realistic residual wavefront errors, photon noise, and thermal background noise.

We found that the star is bright enough that measurements of its flux are dominated by its photon noise, but that flux measurements of the far fainter companion are instead dominated by noise arising from the thermal background (see Figure~\ref{fig:p4_ch5_var_hist}). When we then used this simulation to produce a raw differential light curve of a simulated companion and binned it to the same binning used for the injected companions (Figure~\ref{fig:p4_ch5_combined_sim_planet_variability}), we found that the scatter of the data points is very similar to those of the non-varying artificial companions. Although this simulation is by no means comprehensive and does not include every source of systematics, this suggests that a significant fraction of the RMS measured for the real data in Section~\ref{p4_ch5_discuss_precision} can be accounted for by the photon noise and thermal background. As mentioned previously, the scatter due to both of these effects can be reduced by binning frames. This is therefore consistent with the trends that we see in Figure~\ref{fig:p4_ch5_rms_plot}, which show that increasing the bin size with additional data will continue to improve the precision achieved in differential light curves.
\section{Conclusions}\label{p4_ch5_conclusions}
We present an analysis of the vAPP-enabled ground-based differential spectrophotometry technique for measuring the variability of high-contrast substellar companions, in which we explore the systematics that limit the precision achieved with this technique.

We injected artificial companions with and without simulated variability signals into real observational data at different locations. The data used for this study were the L-band LBT/ALES+dgvAPP360 observations of the HD~1160~system first presented by \citet{2023MNRAS.520.4235S}, who used this technique to measure the variability of substellar companion HD~1160~B. Injecting artificial companions with no variability allowed us to assess the extent to which telluric, instrumental, and other systematics contaminate the data, while artificial time-varying companions let us test how well the injected variability can be recovered. Uniquely for a direct imaging study, we used the instantaneous stellar PSF in each frame as the template for the artificial companion, thus capturing frame-to-frame systematic variations that would affect a real companion. We injected artificial companions at 90\textdegree{} interval offsets from HD~1160~B, with the same 6.35~mag contrast as this companion, and used the methodology presented by \citet{2023MNRAS.520.4235S} to process the data and extract spectrophotometry for the host star and each artificial companion. We used the measured variability of HD~1160~B as the variability signal for the varying artificial companions, such that HD~1160~B could be used as a baseline for comparison. We then produced differential white-light curves for each companion and detrended them using a multiple linear regression approach.

Using Lomb-Scargle periodograms, we find that we successfully recover the injected variability signal to a high significance for the time-varying companions at 90\textdegree{} and 180\textdegree{} offsets from the real companion, and do not find any significant peaks for the non-varying companions injected at these positions. However, we highlight that while our use of the instantaneous stellar PSF as the companion injection template improves on the literature approach, the caveat that this does not account for systematics arising from differences in colour between the star and companion remains. Systematics related to this colour difference may therefore lead to an increased recovery significance in some cases. Furthermore, the periodograms for both the varying and non-varying companions injected at a 270\textdegree{} offset contain multiple peaks with false-alarm probabilities smaller than 1\%, suggesting that companions at this location suffer from periodic systematics. Potential causes could include field-dependent
aberrations such as detector bad pixels or stray light, in the case that these are only partially mitigated by the detrending process. Alternative, more sophisticated detrending approaches may help to further correct for such effects in future studies.

When fitting the variability, we also find that the properties of the recovered sinusoids (amplitude, period, and phase) do not perfectly match the injected sinusoid, indicating that a single night of data ($\sim$7.81~h) is insufficient to accurately measure variability properties for variability of a 3.24~h period (i.e. $\sim$2.41 periods are obtained). Observations with a longer baseline covering a larger number of periods are therefore required to obtain accurate measurements of companion variability properties.

We find that the RMS of the detrended differential white-light curves of the injected companions decreases with increasing bin size according to the white noise model without plateauing, consistent with the result found by \citet{2023MNRAS.520.4235S} for their light curve of HD~1160~B. This suggests that the data is not systematic-limited and that additional data could allow a greater precision to be reached. We also find that the RMS measurements of the artificial companions injected with variability are generally higher than those without variability. The RMS values for the light curves of the non-varying companions range from 0.0492-0.0696 for a binning of 200 frames per bin, which, when interpreted as the sensitivity limit in each case, suggest a limiting precision of 4.9-7.0\% at this bin size. The 8.8\% semi-amplitude variability measured by \citet{2023MNRAS.520.4235S} for HD~1160~B is higher than this level and so we conclude that this remains consistent with astrophysical variability. However, the sensitivity that can be achieved with this technique is inherently dependent on the contrast of the targets; the light curves of fainter companions will have a higher scatter and thus observations will be less sensitive to variability. We conducted an initial investigation into the precision that can be achieved for fainter companions by injecting example artificial companions at contrasts 1 and 2 magnitudes fainter than HD~1160~B. We find precisions at the 14.4\% and 36.6\% level for these injections, respectively, when the light curves are binned to the same bin size. However, we also find that these precision levels can be improved significantly with only a moderate decrease in cadence. In the 1-magnitude fainter case, an increase from 18~minutes to $\sim$40 minutes improves the precision to 5.7\%. Thus, we emphasise the importance of the trade-off between precision and cadence when producing light curves for high-contrast companions, while noting the caveat that it is difficult to draw definitive conclusions about the sensitivity of this technique from a single dataset alone.

We also used simulated data matching the LBT/ALES+dgvAPP360 instrumental setup, produced using the Python package HCIPy \citep{2018SPIE10703E..42P}, to test the effects of specific and known sources of systematics such as NCPAs and AO residuals. HCIPy correctly models coupled effects between NCPAs and AO residuals, propagating them through to the simulated data. First, we tested the impact of NCPAs on star and companion photometry by adding them to the simulated data using 100 Zernike modes. We find that when we scale the wavefront error of these aberrations with a realistic power law, high-order aberrations do not have a significant impact on the fluxes of the targets, but low-order modes can cause a significant reduction in the average measured flux of the companion over the observing sequence. However, we find that the variation in flux over the observing sequence for a given mode is <1\%, suggesting that companion variability measurements may be minimally affected by these aberrations. 

We then simulated realistic residual wavefront errors, photon noise, and thermal background noise, this time neglecting NCPAs due to the very high Strehl ($\sim$98\%) achieved by the LBTI AO system at this wavelength and their very low impact on variability measurements seen in the previous simulations. We showed that flux measurements of the bright host star are dominated by its photon noise, while thermal background noise is the dominant effect for flux measurements of the simulated companion. We used this simulation to produce a detrended differential light curve and found that the scatter on the datapoints closely match those of the non-varying injected artificial companions when binned to the same bin size, suggesting that these same noise sources are the dominant effects in the real data. As the effects of both photon noise and background noise decrease with increasing bin size, this is consistent with the RMS trends that we measured for the injected companions.

For future observations, techniques such as predictive control and focal-plane wavefront sensing can help to further mitigate systematics arising from wavefront aberrations, and may therefore enable a greater precision to be achieved with vAPP-enabled differential spectrophotometry. These precision limits could be further characterised in future studies with an expanded range of datasets and a more comprehensive suite of artificial companion injections, potentially probing a variety of contrasts and variability signals.

\section*{Acknowledgements}\label{p4_ch5_ack}
The authors would like to thank Frank Backs for valuable discussions that improved this work. We also thank our anonymous referee whose comments helped us to improve this work. BJS and BB acknowledge funding by the UK Science and Technology Facilities Council (STFC) grant nos. ST/V000594/1 and UKRI1196. BJS was also supported by the Netherlands Research School for Astronomy (NOVA). JLB and LTP acknowledge funding from the European Research Council (ERC) under the European Union’s Horizon 2020 research and innovation program under grant agreement No 805445. This paper is based on work funded by the United States National Science Foundation (NSF) grants 1608834, 1614320, and 1614492. The research of DD and FS leading to these results has received funding from the European Research Council under ERC Starting Grant agreement 678194 (FALCONER).

We acknowledge the use of the Large Binocular Telescope Interferometer (LBTI) and the support from the LBTI team, specifically from Emily Mailhot, Jared Carlson, Jennifer Power, Phil Hinz, Michael Skrutskie, Travis Barman, Ilya Ilyin, Klaus G. Strassmeier, and Ji Wang. The LBT is an international collaboration among institutions in the United States, Italy and Germany. LBT Corporation partners are: The University of Arizona on behalf of the Arizona Board of Regents; Istituto Nazionale di Astrofisica, Italy; LBT Beteiligungsgesellschaft, Germany, representing the Max-Planck Society, The Leibniz Institute for Astrophysics Potsdam, and Heidelberg University; The Ohio State University, representing OSU, University of Notre Dame, University of Minnesota and University of Virginia. We gratefully acknowledge the use of Native land for our observations. We respectfully acknowledge the University of Arizona is on the land and territories of Indigenous peoples. Today, Arizona is home to 22 federally recognized tribes, with Tucson being home to the O’odham and the Yaqui. Committed to diversity and inclusion, the University strives to build sustainable relationships with sovereign Native Nations and Indigenous communities through education offerings, partnerships, and community service. LBT observations were conducted on the land of the Ndee/Nnēē, Chiricahua, Mescalero, and San Carlos Apache tribes.

This research has made use of NASA’s Astrophysics Data System. This research has made use of the NASA Exoplanet Archive, which is operated by the California Institute of Technology, under contract with the National Aeronautics and Space Administration under the Exoplanet Exploration Program. This research has made use of the SIMBAD database, operated at CDS, Strasbourg, France \citep{2000A&AS..143....9W}. This research made use of SAOImageDS9, a tool for data visualization supported by the Chandra X-ray Science Center (CXC) and the High Energy Astrophysics Science Archive Center (HEASARC) with support from the JWST Mission office at the Space Telescope Science Institute for 3D visualization \citep{2003ASPC..295..489J}. This work made use of the whereistheplanet\footnote{\url{http://whereistheplanet.com/}} prediction tool \citep{2021ascl.soft01003W}. This work makes use of the Python programming language\footnote{Python Software Foundation; \url{https://www.python.org/}}, in particular packages including Matplotlib \citep{Hunter:2007}, NumPy \citep{harris2020array}, SciPy \citep{2020SciPy-NMeth}, Astropy \citep{2013A&A...558A..33A, 2018AJ....156..123A, 2022ApJ...935..167A}, Photutils \citep{larry_bradley_2022_6385735}, scikit-learn \citep{scikit-learn}, statsmodels \citep{seabold2010statsmodels}, pandas \citep{mckinney-proc-scipy-2010, jeff_reback_2022_6408044}, HCIPy \citep{2018SPIE10703E..42P}, PyAstronomy \citep{pya}, and PynPoint \citep{2012MNRAS.427..948A, 2019A&A...621A..59S}.
\section*{Data Availability}\label{p4_ch5_ava}
The data from the HCIPy simulations and LBT/ALES+dgvAPP360 observations underlying this article will be available in the Research Data Management Zenodo repository of the Anton Pannekoek Institute for Astronomy shortly after publication, at \url{https://doi.org/10.5281/zenodo.7603220}.




\bibliographystyle{mnras}
\bibliography{bibliography} 

\begin{thebibliography}{}
\makeatletter
\relax
\def\mn@urlcharsother{\let\do\@makeother \do\$\do\&\do\#\do\^\do\_\do\%\do\~}
\def\mn@doi{\begingroup\mn@urlcharsother \@ifnextchar [ {\mn@doi@}
  {\mn@doi@[]}}
\def\mn@doi@[#1]#2{\def\@tempa{#1}\ifx\@tempa\@empty \href
  {http://dx.doi.org/#2} {doi:#2}\else \href {http://dx.doi.org/#2} {#1}\fi
  \endgroup}
\def\mn@eprint#1#2{\mn@eprint@#1:#2::\@nil}
\def\mn@eprint@arXiv#1{\href {http://arxiv.org/abs/#1} {{\tt arXiv:#1}}}
\def\mn@eprint@dblp#1{\href {http://dblp.uni-trier.de/rec/bibtex/#1.xml}
  {dblp:#1}}
\def\mn@eprint@#1:#2:#3:#4\@nil{\def\@tempa {#1}\def\@tempb {#2}\def\@tempc
  {#3}\ifx \@tempc \@empty \let \@tempc \@tempb \let \@tempb \@tempa \fi \ifx
  \@tempb \@empty \def\@tempb {arXiv}\fi \@ifundefined
  {mn@eprint@\@tempb}{\@tempb:\@tempc}{\expandafter \expandafter \csname
  mn@eprint@\@tempb\endcsname \expandafter{\@tempc}}}

\bibitem[\protect\citeauthoryear{{Amara} \& {Quanz}}{{Amara} \&
  {Quanz}}{2012}]{2012MNRAS.427..948A}
{Amara} A.,  {Quanz} S.~P.,  2012, \mn@doi [\mnras]
  {10.1111/j.1365-2966.2012.21918.x}, \href
  {https://ui.adsabs.harvard.edu/abs/2012MNRAS.427..948A} {427, 948}

\bibitem[\protect\citeauthoryear{{Apai}, {Radigan}, {Buenzli}, {Burrows},
  {Reid}  \& {Jayawardhana}}{{Apai} et~al.}{2013}]{2013ApJ...768..121A}
{Apai} D.,  {Radigan} J.,  {Buenzli} E.,  {Burrows} A.,  {Reid} I.~N.,
  {Jayawardhana} R.,  2013, \mn@doi [\apj] {10.1088/0004-637X/768/2/121}, \href
  {https://ui.adsabs.harvard.edu/abs/2013ApJ...768..121A} {768, 121}

\bibitem[\protect\citeauthoryear{{Apai} et~al.,}{{Apai}
  et~al.}{2016}]{2016ApJ...820...40A}
{Apai} D.,  et~al., 2016, \mn@doi [\apj] {10.3847/0004-637X/820/1/40}, \href
  {https://ui.adsabs.harvard.edu/abs/2016ApJ...820...40A} {820, 40}

\bibitem[\protect\citeauthoryear{{Apai} et~al.,}{{Apai}
  et~al.}{2017}]{2017Sci...357..683A}
{Apai} D.,  et~al., 2017, \mn@doi [Science] {10.1126/science.aam9848}, \href
  {https://ui.adsabs.harvard.edu/abs/2017Sci...357..683A} {357, 683}

\bibitem[\protect\citeauthoryear{{Artigau}, {Bouchard}, {Doyon}  \&
  {Lafreni{\`e}re}}{{Artigau} et~al.}{2009}]{2009ApJ...701.1534A}
{Artigau} {\'E}.,  {Bouchard} S.,  {Doyon} R.,   {Lafreni{\`e}re} D.,  2009,
  \mn@doi [\apj] {10.1088/0004-637X/701/2/1534}, \href
  {https://ui.adsabs.harvard.edu/abs/2009ApJ...701.1534A} {701, 1534}

\bibitem[\protect\citeauthoryear{{Astropy Collaboration} et~al.,}{{Astropy
  Collaboration} et~al.}{2013}]{2013A&A...558A..33A}
{Astropy Collaboration} et~al., 2013, \mn@doi [\aap]
  {10.1051/0004-6361/201322068}, \href
  {https://ui.adsabs.harvard.edu/abs/2013A&A...558A..33A} {558, A33}

\bibitem[\protect\citeauthoryear{{Astropy Collaboration} et~al.,}{{Astropy
  Collaboration} et~al.}{2018}]{2018AJ....156..123A}
{Astropy Collaboration} et~al., 2018, \mn@doi [\aj] {10.3847/1538-3881/aabc4f},
  \href {https://ui.adsabs.harvard.edu/abs/2018AJ....156..123A} {156, 123}

\bibitem[\protect\citeauthoryear{{Astropy Collaboration} et~al.,}{{Astropy
  Collaboration} et~al.}{2022}]{2022ApJ...935..167A}
{Astropy Collaboration} et~al., 2022, \mn@doi [\apj]
  {10.3847/1538-4357/ac7c74}, \href
  {https://ui.adsabs.harvard.edu/abs/2022ApJ...935..167A} {935, 167}

\bibitem[\protect\citeauthoryear{{Biller} et~al.,}{{Biller}
  et~al.}{2013}]{2013ApJ...778L..10B}
{Biller} B.~A.,  et~al., 2013, \mn@doi [\apjl] {10.1088/2041-8205/778/1/L10},
  \href {https://ui.adsabs.harvard.edu/abs/2013ApJ...778L..10B} {778, L10}

\bibitem[\protect\citeauthoryear{{Biller} et~al.,}{{Biller}
  et~al.}{2015}]{2015ApJ...813L..23B}
{Biller} B.~A.,  et~al., 2015, \mn@doi [\apjl] {10.1088/2041-8205/813/2/L23},
  \href {https://ui.adsabs.harvard.edu/abs/2015ApJ...813L..23B} {813, L23}

\bibitem[\protect\citeauthoryear{{Biller} et~al.,}{{Biller}
  et~al.}{2018}]{2018AJ....155...95B}
{Biller} B.~A.,  et~al., 2018, \mn@doi [\aj] {10.3847/1538-3881/aaa5a6}, \href
  {https://ui.adsabs.harvard.edu/abs/2018AJ....155...95B} {155, 95}

\bibitem[\protect\citeauthoryear{{Biller} et~al.,}{{Biller}
  et~al.}{2021}]{2021MNRAS.503..743B}
{Biller} B.~A.,  et~al., 2021, \mn@doi [\mnras] {10.1093/mnras/stab202}, \href
  {https://ui.adsabs.harvard.edu/abs/2021MNRAS.503..743B} {503, 743}

\bibitem[\protect\citeauthoryear{{Biller} et~al.,}{{Biller}
  et~al.}{2024}]{2024MNRAS.532.2207B}
{Biller} B.~A.,  et~al., 2024, \mn@doi [\mnras] {10.1093/mnras/stae1602}, \href
  {https://ui.adsabs.harvard.edu/abs/2024MNRAS.532.2207B} {532, 2207}

\bibitem[\protect\citeauthoryear{{Bonnefoy} et~al.,}{{Bonnefoy}
  et~al.}{2011}]{2011A&A...528L..15B}
{Bonnefoy} M.,  et~al., 2011, \mn@doi [\aap] {10.1051/0004-6361/201016224},
  \href {https://ui.adsabs.harvard.edu/abs/2011A&A...528L..15B} {528, L15}

\bibitem[\protect\citeauthoryear{{Bos} et~al.,}{{Bos}
  et~al.}{2019}]{2019A&A...632A..48B}
{Bos} S.~P.,  et~al., 2019, \mn@doi [\aap] {10.1051/0004-6361/201936062}, \href
  {https://ui.adsabs.harvard.edu/abs/2019A&A...632A..48B} {632, A48}

\bibitem[\protect\citeauthoryear{{Bos} et~al.,}{{Bos}
  et~al.}{2020}]{2020A&A...639A..52B}
{Bos} S.~P.,  et~al., 2020, \mn@doi [\aap] {10.1051/0004-6361/202037910}, \href
  {https://ui.adsabs.harvard.edu/abs/2020A&A...639A..52B} {639, A52}

\bibitem[\protect\citeauthoryear{{Bowler}, {Zhou}, {Morley}, {Kataria},
  {Bryan}, {Benneke}  \& {Batygin}}{{Bowler}
  et~al.}{2020}]{2020ApJ...893L..30B}
{Bowler} B.~P.,  {Zhou} Y.,  {Morley} C.~V.,  {Kataria} T.,  {Bryan} M.~L.,
  {Benneke} B.,   {Batygin} K.,  2020, \mn@doi [\apjl]
  {10.3847/2041-8213/ab8197}, \href
  {https://ui.adsabs.harvard.edu/abs/2020ApJ...893L..30B} {893, L30}

\bibitem[\protect\citeauthoryear{Bradley et~al.,}{Bradley
  et~al.}{2022}]{larry_bradley_2022_6385735}
Bradley L.,  et~al., 2022, astropy/photutils:, \mn@doi{10.5281/zenodo.6385735},
  \url {https://doi.org/10.5281/zenodo.6385735}

\bibitem[\protect\citeauthoryear{{Briesemeister} et~al.,}{{Briesemeister}
  et~al.}{2018}]{2018SPIE10702E..2QB}
{Briesemeister} Z.,  et~al., 2018, in {Evans} C.~J.,  {Simard} L.,   {Takami}
  H.,  eds,  Society of Photo-Optical Instrumentation Engineers (SPIE)
  Conference Series Vol. 10702, Ground-based and Airborne Instrumentation for
  Astronomy VII. p. 107022Q (\mn@eprint {arXiv} {1808.03356}),
  \mn@doi{10.1117/12.2312859}

\bibitem[\protect\citeauthoryear{{Briesemeister} et~al.,}{{Briesemeister}
  et~al.}{2019}]{2019AJ....157..244B}
{Briesemeister} Z.~W.,  et~al., 2019, \mn@doi [\aj] {10.3847/1538-3881/ab1901},
  \href {https://ui.adsabs.harvard.edu/abs/2019AJ....157..244B} {157, 244}

\bibitem[\protect\citeauthoryear{{Broeg}, {Fern{\'a}ndez}  \&
  {Neuh{\"a}user}}{{Broeg} et~al.}{2005}]{2005AN....326..134B}
{Broeg} C.,  {Fern{\'a}ndez} M.,   {Neuh{\"a}user} R.,  2005, \mn@doi
  [Astronomische Nachrichten] {10.1002/asna.200410350}, \href
  {https://ui.adsabs.harvard.edu/abs/2005AN....326..134B} {326, 134}

\bibitem[\protect\citeauthoryear{{Cantalloube} et~al.,}{{Cantalloube}
  et~al.}{2018}]{2018A&A...620L..10C}
{Cantalloube} F.,  et~al., 2018, \mn@doi [\aap] {10.1051/0004-6361/201834311},
  \href {https://ui.adsabs.harvard.edu/abs/2018A&A...620L..10C} {620, L10}

\bibitem[\protect\citeauthoryear{{Cantalloube} et~al.,}{{Cantalloube}
  et~al.}{2020}]{2020A&A...638A..98C}
{Cantalloube} F.,  et~al., 2020, \mn@doi [\aap] {10.1051/0004-6361/201937397},
  \href {https://ui.adsabs.harvard.edu/abs/2020A&A...638A..98C} {638, A98}

\bibitem[\protect\citeauthoryear{{Chen} et~al.,}{{Chen}
  et~al.}{2025}]{2025MNRAS.539.3758C}
{Chen} X.,  et~al., 2025, \mn@doi [\mnras] {10.1093/mnras/staf737}, \href
  {https://ui.adsabs.harvard.edu/abs/2025MNRAS.539.3758C} {539, 3758}

\bibitem[\protect\citeauthoryear{{Cushing} et~al.,}{{Cushing}
  et~al.}{2016}]{2016ApJ...823..152C}
{Cushing} M.~C.,  et~al., 2016, \mn@doi [\apj] {10.3847/0004-637X/823/2/152},
  \href {https://ui.adsabs.harvard.edu/abs/2016ApJ...823..152C} {823, 152}

\bibitem[\protect\citeauthoryear{{Czesla}, {Schr{\"o}ter}, {Schneider},
  {Huber}, {Pfeifer}, {Andreasen}  \& {Zechmeister}}{{Czesla}
  et~al.}{2019}]{pya}
{Czesla} S.,  {Schr{\"o}ter} S.,  {Schneider} C.~P.,  {Huber} K.~F.,  {Pfeifer}
  F.,  {Andreasen} D.~T.,   {Zechmeister} M.,  2019, {PyA: Python
  astronomy-related packages} (\mn@eprint {ascl} {1906.010})

\bibitem[\protect\citeauthoryear{{Diamond-Lowe}, {Berta-Thompson},
  {Charbonneau}  \& {Kempton}}{{Diamond-Lowe}
  et~al.}{2018}]{2018AJ....156...42D}
{Diamond-Lowe} H.,  {Berta-Thompson} Z.,  {Charbonneau} D.,   {Kempton} E.
  M.~R.,  2018, \mn@doi [\aj] {10.3847/1538-3881/aac6dd}, \href
  {https://ui.adsabs.harvard.edu/abs/2018AJ....156...42D} {156, 42}

\bibitem[\protect\citeauthoryear{{Diamond-Lowe}, {Mendon{\c{c}}a},
  {Charbonneau}  \& {Buchhave}}{{Diamond-Lowe}
  et~al.}{2023}]{2023AJ....165..169D}
{Diamond-Lowe} H.,  {Mendon{\c{c}}a} J.~M.,  {Charbonneau} D.,   {Buchhave}
  L.~A.,  2023, \mn@doi [\aj] {10.3847/1538-3881/acbf39}, \href
  {https://ui.adsabs.harvard.edu/abs/2023AJ....165..169D} {165, 169}

\bibitem[\protect\citeauthoryear{{Doelman}, {Snik}, {Warriner}  \&
  {Escuti}}{{Doelman} et~al.}{2017}]{2017SPIE10400E..0UD}
{Doelman} D.~S.,  {Snik} F.,  {Warriner} N.~Z.,   {Escuti} M.~J.,  2017, in
  {Shaklan} S.,  ed.,  Society of Photo-Optical Instrumentation Engineers
  (SPIE) Conference Series Vol. 10400, Society of Photo-Optical Instrumentation
  Engineers (SPIE) Conference Series. p. 104000U (\mn@eprint {arXiv}
  {1709.09897}), \mn@doi{10.1117/12.2273406}

\bibitem[\protect\citeauthoryear{{Doelman}, {Por}, {Ruane}, {Escuti}  \&
  {Snik}}{{Doelman} et~al.}{2020}]{2020PASP..132d5002D}
{Doelman} D.~S.,  {Por} E.~H.,  {Ruane} G.,  {Escuti} M.~J.,   {Snik} F.,
  2020, \mn@doi [\pasp] {10.1088/1538-3873/ab755f}, \href
  {https://ui.adsabs.harvard.edu/abs/2020PASP..132d5002D} {132, 045002}

\bibitem[\protect\citeauthoryear{{Doelman} et~al.,}{{Doelman}
  et~al.}{2021}]{2021ApOpt..60D..52D}
{Doelman} D.~S.,  et~al., 2021, \mn@doi [\ao] {10.1364/AO.422155}, \href
  {https://ui.adsabs.harvard.edu/abs/2021ApOpt..60D..52D} {60, D52}

\bibitem[\protect\citeauthoryear{{Doelman} et~al.,}{{Doelman}
  et~al.}{2022}]{2022AJ....163..217D}
{Doelman} D.~S.,  et~al., 2022, \mn@doi [\aj] {10.3847/1538-3881/ac5d52}, \href
  {https://ui.adsabs.harvard.edu/abs/2022AJ....163..217D} {163, 217}

\bibitem[\protect\citeauthoryear{{Ertel} et~al.,}{{Ertel}
  et~al.}{2020}]{2020SPIE11446E..07E}
{Ertel} S.,  et~al., 2020, in Society of Photo-Optical Instrumentation
  Engineers (SPIE) Conference Series. p. 1144607, \mn@doi{10.1117/12.2561849}

\bibitem[\protect\citeauthoryear{{Fowler}, {Van Kooten}  \&
  {Jensen-Clem}}{{Fowler} et~al.}{2022}]{2022SPIE12185E..82F}
{Fowler} J.,  {Van Kooten} M.~A.~M.,   {Jensen-Clem} R.,  2022, in {Schreiber}
  L.,  {Schmidt} D.,   {Vernet} E.,  eds,  Society of Photo-Optical
  Instrumentation Engineers (SPIE) Conference Series Vol. 12185, Adaptive
  Optics Systems VIII. p. 1218582 (\mn@eprint {arXiv} {2208.00984}),
  \mn@doi{10.1117/12.2629521}

\bibitem[\protect\citeauthoryear{{Fuda}, {Apai}, {Nardiello}, {Tan}, {Karalidi}
   \& {Bedin}}{{Fuda} et~al.}{2024}]{2024ApJ...965..182F}
{Fuda} N.,  {Apai} D.,  {Nardiello} D.,  {Tan} X.,  {Karalidi} T.,   {Bedin}
  L.~R.,  2024, \mn@doi [\apj] {10.3847/1538-4357/ad2c84}, \href
  {https://ui.adsabs.harvard.edu/abs/2024ApJ...965..182F} {965, 182}

\bibitem[\protect\citeauthoryear{{Gibson}, {Aigrain}, {Roberts}, {Evans},
  {Osborne}  \& {Pont}}{{Gibson} et~al.}{2012}]{2012MNRAS.419.2683G}
{Gibson} N.~P.,  {Aigrain} S.,  {Roberts} S.,  {Evans} T.~M.,  {Osborne} M.,
  {Pont} F.,  2012, \mn@doi [\mnras] {10.1111/j.1365-2966.2011.19915.x}, \href
  {https://ui.adsabs.harvard.edu/abs/2012MNRAS.419.2683G} {419, 2683}

\bibitem[\protect\citeauthoryear{{Gonsalves}}{{Gonsalves}}{1982}]{1982OptEn..21..829G}
{Gonsalves} R.~A.,  1982, \mn@doi [Optical Engineering] {10.1117/12.7972989},
  \href {https://ui.adsabs.harvard.edu/abs/1982OptEn..21..829G} {21, 829}

\bibitem[\protect\citeauthoryear{{Goulding} et~al.,}{{Goulding}
  et~al.}{2012}]{2012MNRAS.427.3358G}
{Goulding} N.~T.,  et~al., 2012, \mn@doi [\mnras]
  {10.1111/j.1365-2966.2012.21932.x}, \href
  {https://ui.adsabs.harvard.edu/abs/2012MNRAS.427.3358G} {427, 3358}

\bibitem[\protect\citeauthoryear{{Guyon} \& {Males}}{{Guyon} \&
  {Males}}{2017}]{2017arXiv170700570G}
{Guyon} O.,  {Males} J.,  2017, \mn@doi [arXiv e-prints]
  {10.48550/arXiv.1707.00570}, \href
  {https://ui.adsabs.harvard.edu/abs/2017arXiv170700570G} {p. arXiv:1707.00570}

\bibitem[\protect\citeauthoryear{{Haffert} et~al.,}{{Haffert}
  et~al.}{2021}]{2021JATIS...7b9001H}
{Haffert} S.~Y.,  et~al., 2021, \mn@doi [Journal of Astronomical Telescopes,
  Instruments, and Systems] {10.1117/1.JATIS.7.2.029001}, \href
  {https://ui.adsabs.harvard.edu/abs/2021JATIS...7b9001H} {7, 029001}

\bibitem[\protect\citeauthoryear{{Hallinan} et~al.,}{{Hallinan}
  et~al.}{2015}]{2015Natur.523..568H}
{Hallinan} G.,  et~al., 2015, \mn@doi [\nat] {10.1038/nature14619}, \href
  {https://ui.adsabs.harvard.edu/abs/2015Natur.523..568H} {523, 568}

\bibitem[\protect\citeauthoryear{Harris et~al.,}{Harris
  et~al.}{2020}]{harris2020array}
Harris C.~R.,  et~al., 2020, \mn@doi [Nature] {10.1038/s41586-020-2649-2}, 585,
  357

\bibitem[\protect\citeauthoryear{{Hartung} et~al.,}{{Hartung}
  et~al.}{2014}]{2014SPIE.9148E..5QH}
{Hartung} M.,  et~al., 2014, in {Marchetti} E.,  {Close} L.~M.,   {Vran} J.-P.,
   eds,  Society of Photo-Optical Instrumentation Engineers (SPIE) Conference
  Series Vol. 9148, Adaptive Optics Systems IV. p. 91485Q (\mn@eprint {arXiv}
  {1407.7895}), \mn@doi{10.1117/12.2056661}

\bibitem[\protect\citeauthoryear{{Hinkley} et~al.,}{{Hinkley}
  et~al.}{2007}]{2007ApJ...654..633H}
{Hinkley} S.,  et~al., 2007, \mn@doi [\apj] {10.1086/509063}, \href
  {https://ui.adsabs.harvard.edu/abs/2007ApJ...654..633H} {654, 633}

\bibitem[\protect\citeauthoryear{{Hinz} et~al.,}{{Hinz}
  et~al.}{2016}]{2016SPIE.9907E..04H}
{Hinz} P.~M.,  et~al., 2016, in {Malbet} F.,  {Creech-Eakman} M.~J.,
  {Tuthill} P.~G.,  eds,  Society of Photo-Optical Instrumentation Engineers
  (SPIE) Conference Series Vol. 9907, Optical and Infrared Interferometry and
  Imaging V. p. 990704, \mn@doi{10.1117/12.2233795}

\bibitem[\protect\citeauthoryear{{Hinz}, {Skemer}, {Stone}, {Montoya}  \&
  {Durney}}{{Hinz} et~al.}{2018}]{2018SPIE10702E..3LH}
{Hinz} P.~M.,  {Skemer} A.,  {Stone} J.,  {Montoya} O.~M.,   {Durney} O.,
  2018, in {Evans} C.~J.,  {Simard} L.,   {Takami} H.,  eds,  Society of
  Photo-Optical Instrumentation Engineers (SPIE) Conference Series Vol. 10702,
  Ground-based and Airborne Instrumentation for Astronomy VII. p. 107023L,
  \mn@doi{10.1117/12.2314289}

\bibitem[\protect\citeauthoryear{{Horne}}{{Horne}}{1986}]{1986PASP...98..609H}
{Horne} K.,  1986, \mn@doi [\pasp] {10.1086/131801}, \href
  {https://ui.adsabs.harvard.edu/abs/1986PASP...98..609H} {98, 609}

\bibitem[\protect\citeauthoryear{{Horne} \& {Baliunas}}{{Horne} \&
  {Baliunas}}{1986}]{1986ApJ...302..757H}
{Horne} J.~H.,  {Baliunas} S.~L.,  1986, \mn@doi [\apj] {10.1086/164037}, \href
  {https://ui.adsabs.harvard.edu/abs/1986ApJ...302..757H} {302, 757}

\bibitem[\protect\citeauthoryear{Hunter}{Hunter}{2007}]{Hunter:2007}
Hunter J.~D.,  2007, \mn@doi [Computing in Science \& Engineering]
  {10.1109/MCSE.2007.55}, 9, 90

\bibitem[\protect\citeauthoryear{{Isbell} et~al.,}{{Isbell}
  et~al.}{2024}]{2024SPIE13095E..06I}
{Isbell} J.~W.,  et~al., 2024, in {Kammerer} J.,  {Sallum} S.,
  {Sanchez-Bermudez} J.,  eds,  Society of Photo-Optical Instrumentation
  Engineers (SPIE) Conference Series Vol. 13095, Optical and Infrared
  Interferometry and Imaging IX. p. 1309506, \mn@doi{10.1117/12.3027270}

\bibitem[\protect\citeauthoryear{{Jensen-Clem}, {Bond}, {Cetre}, {McEwen},
  {Wizinowich}, {Ragland}, {Mawet}  \& {Graham}}{{Jensen-Clem}
  et~al.}{2019}]{2019SPIE11117E..0WJ}
{Jensen-Clem} R.,  {Bond} C.~Z.,  {Cetre} S.,  {McEwen} E.,  {Wizinowich} P.,
  {Ragland} S.,  {Mawet} D.,   {Graham} J.,  2019, in Society of Photo-Optical
  Instrumentation Engineers (SPIE) Conference Series. p. 111170W (\mn@eprint
  {arXiv} {1909.05302}), \mn@doi{10.1117/12.2529687}

\bibitem[\protect\citeauthoryear{{Joye} \& {Mandel}}{{Joye} \&
  {Mandel}}{2003}]{2003ASPC..295..489J}
{Joye} W.~A.,  {Mandel} E.,  2003, in {Payne} H.~E.,  {Jedrzejewski} R.~I.,
  {Hook} R.~N.,  eds,  Astronomical Society of the Pacific Conference Series
  Vol. 295, Astronomical Data Analysis Software and Systems XII. p.~489

\bibitem[\protect\citeauthoryear{{Karalidi}, {Apai}, {Marley}  \&
  {Buenzli}}{{Karalidi} et~al.}{2016}]{2016ApJ...825...90K}
{Karalidi} T.,  {Apai} D.,  {Marley} M.~S.,   {Buenzli} E.,  2016, \mn@doi
  [\apj] {10.3847/0004-637X/825/2/90}, \href
  {https://ui.adsabs.harvard.edu/abs/2016ApJ...825...90K} {825, 90}

\bibitem[\protect\citeauthoryear{{Kenworthy} \& {Haffert}}{{Kenworthy} \&
  {Haffert}}{2025}]{2025ARA&A..63..179K}
{Kenworthy} M.~A.,  {Haffert} S.~Y.,  2025, \mn@doi [\araa]
  {10.1146/annurev-astro-021225-022840}, \href
  {https://ui.adsabs.harvard.edu/abs/2025ARA&A..63..179K} {63, 179}

\bibitem[\protect\citeauthoryear{{Kipping} \& {Bakos}}{{Kipping} \&
  {Bakos}}{2011}]{2011ApJ...733...36K}
{Kipping} D.,  {Bakos} G.,  2011, \mn@doi [\apj] {10.1088/0004-637X/733/1/36},
  \href {https://ui.adsabs.harvard.edu/abs/2011ApJ...733...36K} {733, 36}

\bibitem[\protect\citeauthoryear{{Lagrange} et~al.,}{{Lagrange}
  et~al.}{2010}]{2010Sci...329...57L}
{Lagrange} A.~M.,  et~al., 2010, \mn@doi [Science] {10.1126/science.1187187},
  \href {https://ui.adsabs.harvard.edu/abs/2010Sci...329...57L} {329, 57}

\bibitem[\protect\citeauthoryear{{Lamb}, {Norton}, {Macintosh}, {Correia},
  {V{\'e}ran}, {Marois}  \& {Sivanandam}}{{Lamb}
  et~al.}{2018}]{2018SPIE10703E..5ML}
{Lamb} M.,  {Norton} A.,  {Macintosh} B.,  {Correia} C.,  {V{\'e}ran} J.-P.,
  {Marois} C.,   {Sivanandam} S.,  2018, in {Close} L.~M.,  {Schreiber} L.,
  {Schmidt} D.,  eds,  Society of Photo-Optical Instrumentation Engineers
  (SPIE) Conference Series Vol. 10703, Adaptive Optics Systems VI. p. 107035M,
  \mn@doi{10.1117/12.2313458}

\bibitem[\protect\citeauthoryear{{Leisenring} et~al.,}{{Leisenring}
  et~al.}{2012}]{2012SPIE.8446E..4FL}
{Leisenring} J.~M.,  et~al., 2012, in {McLean} I.~S.,  {Ramsay} S.~K.,
  {Takami} H.,  eds,  Society of Photo-Optical Instrumentation Engineers (SPIE)
  Conference Series Vol. 8446, Ground-based and Airborne Instrumentation for
  Astronomy IV. p. 84464F, \mn@doi{10.1117/12.924814}

\bibitem[\protect\citeauthoryear{{Lew} et~al.,}{{Lew}
  et~al.}{2020a}]{2020AJ....159..125L}
{Lew} B. W.~P.,  et~al., 2020a, \mn@doi [\aj] {10.3847/1538-3881/ab5f59}, \href
  {https://ui.adsabs.harvard.edu/abs/2020AJ....159..125L} {159, 125}

\bibitem[\protect\citeauthoryear{{Lew} et~al.,}{{Lew}
  et~al.}{2020b}]{2020ApJ...903...15L}
{Lew} B. W.~P.,  et~al., 2020b, \mn@doi [\apj] {10.3847/1538-4357/abb81d},
  \href {https://ui.adsabs.harvard.edu/abs/2020ApJ...903...15L} {903, 15}

\bibitem[\protect\citeauthoryear{{Liu} et~al.,}{{Liu}
  et~al.}{2023}]{2023A&A...674A.115L}
{Liu} P.,  et~al., 2023, \mn@doi [\aap] {10.1051/0004-6361/202245333}, \href
  {https://ui.adsabs.harvard.edu/abs/2023A&A...674A.115L} {674, A115}

\bibitem[\protect\citeauthoryear{{Liu} et~al.,}{{Liu}
  et~al.}{2024}]{2024MNRAS.527.6624L}
{Liu} P.,  et~al., 2024, \mn@doi [\mnras] {10.1093/mnras/stad3502}, \href
  {https://ui.adsabs.harvard.edu/abs/2024MNRAS.527.6624L} {527, 6624}

\bibitem[\protect\citeauthoryear{{Lomb}}{{Lomb}}{1976}]{1976Ap&SS..39..447L}
{Lomb} N.~R.,  1976, \mn@doi [\apss] {10.1007/BF00648343}, \href
  {https://ui.adsabs.harvard.edu/abs/1976Ap&SS..39..447L} {39, 447}

\bibitem[\protect\citeauthoryear{{Long} et~al.,}{{Long}
  et~al.}{2023}]{2023AJ....165..216L}
{Long} J.~D.,  et~al., 2023, \mn@doi [\aj] {10.3847/1538-3881/acbd4b}, \href
  {https://ui.adsabs.harvard.edu/abs/2023AJ....165..216L} {165, 216}

\bibitem[\protect\citeauthoryear{{Madurowicz} et~al.,}{{Madurowicz}
  et~al.}{2019}]{2019JATIS...5d9003M}
{Madurowicz} A.,  et~al., 2019, \mn@doi [Journal of Astronomical Telescopes,
  Instruments, and Systems] {10.1117/1.JATIS.5.4.049003}, \href
  {https://ui.adsabs.harvard.edu/abs/2019JATIS...5d9003M} {5, 049003}

\bibitem[\protect\citeauthoryear{{Maio} et~al.,}{{Maio}
  et~al.}{2025}]{2025A&A...698A..52M}
{Maio} F.,  et~al., 2025, \mn@doi [\aap] {10.1051/0004-6361/202553672}, \href
  {https://ui.adsabs.harvard.edu/abs/2025A&A...698A..52M} {698, A52}

\bibitem[\protect\citeauthoryear{{Males} et~al.,}{{Males}
  et~al.}{2016}]{2016SPIE.9909E..52M}
{Males} J.~R.,  et~al., 2016, in {Marchetti} E.,  {Close} L.~M.,   {V{\'e}ran}
  J.-P.,  eds,  Society of Photo-Optical Instrumentation Engineers (SPIE)
  Conference Series Vol. 9909, Adaptive Optics Systems V. p. 990952,
  \mn@doi{10.1117/12.2234105}

\bibitem[\protect\citeauthoryear{{Males}, {Fitzgerald}, {Belikov}  \&
  {Guyon}}{{Males} et~al.}{2021}]{2021PASP..133j4504M}
{Males} J.~R.,  {Fitzgerald} M.~P.,  {Belikov} R.,   {Guyon} O.,  2021, \mn@doi
  [\pasp] {10.1088/1538-3873/ac0f0c}, \href
  {https://ui.adsabs.harvard.edu/abs/2021PASP..133j4504M} {133, 104504}

\bibitem[\protect\citeauthoryear{{Manjavacas} et~al.,}{{Manjavacas}
  et~al.}{2019a}]{2019AJ....157..101M}
{Manjavacas} E.,  et~al., 2019a, \mn@doi [\aj] {10.3847/1538-3881/aaf88f},
  \href {https://ui.adsabs.harvard.edu/abs/2019AJ....157..101M} {157, 101}

\bibitem[\protect\citeauthoryear{{Manjavacas} et~al.,}{{Manjavacas}
  et~al.}{2019b}]{2019ApJ...875L..15M}
{Manjavacas} E.,  et~al., 2019b, \mn@doi [\apjl] {10.3847/2041-8213/ab13b9},
  \href {https://ui.adsabs.harvard.edu/abs/2019ApJ...875L..15M} {875, L15}

\bibitem[\protect\citeauthoryear{{Manjavacas}, {Karalidi}, {Vos}, {Biller}  \&
  {Lew}}{{Manjavacas} et~al.}{2021}]{2021AJ....162..179M}
{Manjavacas} E.,  {Karalidi} T.,  {Vos} J.~M.,  {Biller} B.~A.,   {Lew} B.
  W.~P.,  2021, \mn@doi [\aj] {10.3847/1538-3881/ac174c}, \href
  {https://ui.adsabs.harvard.edu/abs/2021AJ....162..179M} {162, 179}

\bibitem[\protect\citeauthoryear{{Marois}, {Macintosh}  \&
  {V{\'e}ran}}{{Marois} et~al.}{2010}]{2010SPIE.7736E..1JM}
{Marois} C.,  {Macintosh} B.,   {V{\'e}ran} J.-P.,  2010, in {Ellerbroek}
  B.~L.,  {Hart} M.,  {Hubin} N.,   {Wizinowich} P.~L.,  eds,  Society of
  Photo-Optical Instrumentation Engineers (SPIE) Conference Series Vol. 7736,
  Adaptive Optics Systems II. p. 77361J, \mn@doi{10.1117/12.857225}

\bibitem[\protect\citeauthoryear{{Martinez}, {Loose}, {Aller Carpentier}  \&
  {Kasper}}{{Martinez} et~al.}{2012}]{2012A&A...541A.136M}
{Martinez} P.,  {Loose} C.,  {Aller Carpentier} E.,   {Kasper} M.,  2012,
  \mn@doi [\aap] {10.1051/0004-6361/201118459}, \href
  {https://ui.adsabs.harvard.edu/abs/2012A&A...541A.136M} {541, A136}

\bibitem[\protect\citeauthoryear{{Martinez}, {Kasper}, {Costille}, {Sauvage},
  {Dohlen}, {Puget}  \& {Beuzit}}{{Martinez}
  et~al.}{2013}]{2013A&A...554A..41M}
{Martinez} P.,  {Kasper} M.,  {Costille} A.,  {Sauvage} J.~F.,  {Dohlen} K.,
  {Puget} P.,   {Beuzit} J.~L.,  2013, \mn@doi [\aap]
  {10.1051/0004-6361/201220820}, \href
  {https://ui.adsabs.harvard.edu/abs/2013A&A...554A..41M} {554, A41}

\bibitem[\protect\citeauthoryear{{Mawet} et~al.,}{{Mawet}
  et~al.}{2012}]{2012SPIE.8442E..04M}
{Mawet} D.,  et~al., 2012, in {Clampin} M.~C.,  {Fazio} G.~G.,  {MacEwen}
  H.~A.,   {Oschmann} Jacobus~M. J.,  eds,  Society of Photo-Optical
  Instrumentation Engineers (SPIE) Conference Series Vol. 8442, Space
  Telescopes and Instrumentation 2012: Optical, Infrared, and Millimeter Wave.
  p. 844204 (\mn@eprint {arXiv} {1207.5481}), \mn@doi{10.1117/12.927245}

\bibitem[\protect\citeauthoryear{{McCarthy} et~al.,}{{McCarthy}
  et~al.}{2024}]{2024ApJ...965...83M}
{McCarthy} A.~M.,  et~al., 2024, \mn@doi [\apj] {10.3847/1538-4357/ad2c76},
  \href {https://ui.adsabs.harvard.edu/abs/2024ApJ...965...83M} {965, 83}

\bibitem[\protect\citeauthoryear{{McCarthy} et~al.,}{{McCarthy}
  et~al.}{2025}]{2025ApJ...981L..22M}
{McCarthy} A.~M.,  et~al., 2025, \mn@doi [\apjl] {10.3847/2041-8213/ad9eaf},
  \href {https://ui.adsabs.harvard.edu/abs/2025ApJ...981L..22M} {981, L22}

\bibitem[\protect\citeauthoryear{{M}c{K}inney}{{M}c{K}inney}{2010}]{mckinney-proc-scipy-2010}
{M}c{K}inney W.,  2010, in {S}t\'efan van~der {W}alt {J}arrod {M}illman eds,
  {P}roceedings of the 9th {P}ython in {S}cience {C}onference. pp 56 -- 61,
  \mn@doi{10.25080/Majora-92bf1922-00a}

\bibitem[\protect\citeauthoryear{{Mendui{\~n}a-Fern{\'a}ndez}, {Tecza}  \&
  {Thatte}}{{Mendui{\~n}a-Fern{\'a}ndez} et~al.}{2020}]{2020SPIE11447E..2LM}
{Mendui{\~n}a-Fern{\'a}ndez} {\'A}.,  {Tecza} M.,   {Thatte} N.,  2020, in
  Society of Photo-Optical Instrumentation Engineers (SPIE) Conference Series.
  p. 114472L, \mn@doi{10.1117/12.2560994}

\bibitem[\protect\citeauthoryear{{Metchev} et~al.,}{{Metchev}
  et~al.}{2015}]{2015ApJ...799..154M}
{Metchev} S.~A.,  et~al., 2015, \mn@doi [\apj] {10.1088/0004-637X/799/2/154},
  \href {https://ui.adsabs.harvard.edu/abs/2015ApJ...799..154M} {799, 154}

\bibitem[\protect\citeauthoryear{{Miles-P{\'a}ez} et~al.,}{{Miles-P{\'a}ez}
  et~al.}{2019}]{2019ApJ...883..181M}
{Miles-P{\'a}ez} P.~A.,  et~al., 2019, \mn@doi [\apj]
  {10.3847/1538-4357/ab3d25}, \href
  {https://ui.adsabs.harvard.edu/abs/2019ApJ...883..181M} {883, 181}

\bibitem[\protect\citeauthoryear{{Miles-P{\'a}ez}, {Metchev}  \&
  {George}}{{Miles-P{\'a}ez} et~al.}{2023}]{2023MNRAS.521..952M}
{Miles-P{\'a}ez} P.~A.,  {Metchev} S.~A.,   {George} B.,  2023, \mn@doi
  [\mnras] {10.1093/mnras/stad273}, \href
  {https://ui.adsabs.harvard.edu/abs/2023MNRAS.521..952M} {521, 952}

\bibitem[\protect\citeauthoryear{{Miller} et~al.,}{{Miller}
  et~al.}{2018}]{2018SPIE10703E..1TM}
{Miller} K.,  et~al., 2018, in {Close} L.~M.,  {Schreiber} L.,   {Schmidt} D.,
  eds,  Society of Photo-Optical Instrumentation Engineers (SPIE) Conference
  Series Vol. 10703, Adaptive Optics Systems VI. p. 107031T (\mn@eprint {arXiv}
  {1807.04381}), \mn@doi{10.1117/12.2312809}

\bibitem[\protect\citeauthoryear{{N'Diaye}, {Dohlen}, {Fusco}  \&
  {Paul}}{{N'Diaye} et~al.}{2013}]{2013A&A...555A..94N}
{N'Diaye} M.,  {Dohlen} K.,  {Fusco} T.,   {Paul} B.,  2013, \mn@doi [\aap]
  {10.1051/0004-6361/201219797}, \href
  {https://ui.adsabs.harvard.edu/abs/2013A&A...555A..94N} {555, A94}

\bibitem[\protect\citeauthoryear{{N'Diaye} et~al.,}{{N'Diaye}
  et~al.}{2014}]{2014SPIE.9148E..5HN}
{N'Diaye} M.,  et~al., 2014, in {Marchetti} E.,  {Close} L.~M.,   {Vran} J.-P.,
   eds,  Society of Photo-Optical Instrumentation Engineers (SPIE) Conference
  Series Vol. 9148, Adaptive Optics Systems IV. p. 91485H,
  \mn@doi{10.1117/12.2056818}

\bibitem[\protect\citeauthoryear{{Naud}, {Artigau}, {Rowe}, {Doyon}, {Malo},
  {Albert}, {Gagn{\'e}}  \& {Bouchard}}{{Naud}
  et~al.}{2017}]{2017AJ....154..138N}
{Naud} M.-E.,  {Artigau} {\'E}.,  {Rowe} J.~F.,  {Doyon} R.,  {Malo} L.,
  {Albert} L.,  {Gagn{\'e}} J.,   {Bouchard} S.,  2017, \mn@doi [\aj]
  {10.3847/1538-3881/aa83b7}, \href
  {https://ui.adsabs.harvard.edu/abs/2017AJ....154..138N} {154, 138}

\bibitem[\protect\citeauthoryear{{Nielsen} et~al.,}{{Nielsen}
  et~al.}{2012}]{2012ApJ...750...53N}
{Nielsen} E.~L.,  et~al., 2012, \mn@doi [\apj] {10.1088/0004-637X/750/1/53},
  \href {https://ui.adsabs.harvard.edu/abs/2012ApJ...750...53N} {750, 53}

\bibitem[\protect\citeauthoryear{{Niu} \& {Tian}}{{Niu} \&
  {Tian}}{2022}]{2022JOpt...24l3001N}
{Niu} K.,  {Tian} C.,  2022, \mn@doi [Journal of Optics]
  {10.1088/2040-8986/ac9e08}, \href
  {https://ui.adsabs.harvard.edu/abs/2022JOpt...24l3001N} {24, 123001}

\bibitem[\protect\citeauthoryear{{Noll}}{{Noll}}{1976}]{1976JOSA...66..207N}
{Noll} R.~J.,  1976, Journal of the Optical Society of America (1917-1983),
  \href {https://ui.adsabs.harvard.edu/abs/1976JOSA...66..207N} {66, 207}

\bibitem[\protect\citeauthoryear{{Otten}, {Snik}, {Kenworthy}, {Miskiewicz}  \&
  {Escuti}}{{Otten} et~al.}{2014a}]{2014OExpr..2230287O}
{Otten} G. P.~P.~L.,  {Snik} F.,  {Kenworthy} M.~A.,  {Miskiewicz} M.~N.,
  {Escuti} M.~J.,  2014a, \mn@doi [Optics Express] {10.1364/OE.22.030287},
  \href {https://ui.adsabs.harvard.edu/abs/2014OExpr..2230287O} {22, 30287}

\bibitem[\protect\citeauthoryear{{Otten}, Snik, Kenworthy, Miskiewicz, Escuti
  \& Codona}{{Otten} et~al.}{2014b}]{10.1117/12.2056096}
{Otten} G. P.~P.~L.,  Snik F.,  Kenworthy M.~A.,  Miskiewicz M.~N.,  Escuti
  M.~J.,   Codona J.~L.,  2014b, in Navarro R.,  Cunningham C.~R.,   Barto
  A.~A.,  eds,  Society of Photo-Optical Instrumentation Engineers (SPIE)
  Conference Series Vol. 9151, Advances in Optical and Mechanical Technologies
  for Telescopes and Instrumentation. SPIE, pp 577 -- 586,
  \mn@doi{10.1117/12.2056096}, \url {https://doi.org/10.1117/12.2056096}

\bibitem[\protect\citeauthoryear{{Otten} et~al.,}{{Otten}
  et~al.}{2017}]{2017ApJ...834..175O}
{Otten} G. P.~P.~L.,  et~al., 2017, \mn@doi [\apj]
  {10.3847/1538-4357/834/2/175}, \href
  {https://ui.adsabs.harvard.edu/abs/2017ApJ...834..175O} {834, 175}

\bibitem[\protect\citeauthoryear{{Panwar}, {D{\'e}sert}, {Todorov}, {Bean},
  {Stevenson}, {Huitson}, {Fortney}  \& {Bergmann}}{{Panwar}
  et~al.}{2022a}]{2022MNRAS.510.3236P}
{Panwar} V.,  {D{\'e}sert} J.-M.,  {Todorov} K.~O.,  {Bean} J.~L.,  {Stevenson}
  K.~B.,  {Huitson} C.~M.,  {Fortney} J.~J.,   {Bergmann} M.,  2022a, \mn@doi
  [\mnras] {10.1093/mnras/stab3646}, \href
  {https://ui.adsabs.harvard.edu/abs/2022MNRAS.510.3236P} {510, 3236}

\bibitem[\protect\citeauthoryear{{Panwar}, {D{\'e}sert}, {Todorov}, {Bean},
  {Stevenson}, {Huitson}, {Fortney}  \& {Bergmann}}{{Panwar}
  et~al.}{2022b}]{2022MNRAS.515.5018P}
{Panwar} V.,  {D{\'e}sert} J.-M.,  {Todorov} K.~O.,  {Bean} J.~L.,  {Stevenson}
  K.~B.,  {Huitson} C.~M.,  {Fortney} J.~J.,   {Bergmann} M.,  2022b, \mn@doi
  [\mnras] {10.1093/mnras/stac1949}, \href
  {https://ui.adsabs.harvard.edu/abs/2022MNRAS.515.5018P} {515, 5018}

\bibitem[\protect\citeauthoryear{{Paxman}, {Schulz}  \& {Fienup}}{{Paxman}
  et~al.}{1992}]{1992JOSAA...9.1072P}
{Paxman} R.~G.,  {Schulz} T.~J.,   {Fienup} J.~R.,  1992, \mn@doi [Journal of
  the Optical Society of America A] {10.1364/JOSAA.9.001072}, \href
  {https://ui.adsabs.harvard.edu/abs/1992JOSAA...9.1072P} {9, 1072}

\bibitem[\protect\citeauthoryear{Pedregosa et~al.,}{Pedregosa
  et~al.}{2011}]{scikit-learn}
Pedregosa F.,  et~al., 2011, Journal of Machine Learning Research, 12, 2825

\bibitem[\protect\citeauthoryear{{Plummer}, {Wang}, {Artigau}, {Doyon}  \&
  {Su{\'a}rez}}{{Plummer} et~al.}{2024}]{2024ApJ...970...62P}
{Plummer} M.~K.,  {Wang} J.,  {Artigau} {\'E}.,  {Doyon} R.,   {Su{\'a}rez} G.,
   2024, \mn@doi [\apj] {10.3847/1538-4357/ad4f89}, \href
  {https://ui.adsabs.harvard.edu/abs/2024ApJ...970...62P} {970, 62}

\bibitem[\protect\citeauthoryear{{Pont}, {Zucker}  \& {Queloz}}{{Pont}
  et~al.}{2006}]{2006MNRAS.373..231P}
{Pont} F.,  {Zucker} S.,   {Queloz} D.,  2006, \mn@doi [\mnras]
  {10.1111/j.1365-2966.2006.11012.x}, \href
  {https://ui.adsabs.harvard.edu/abs/2006MNRAS.373..231P} {373, 231}

\bibitem[\protect\citeauthoryear{{Por}, {Haffert}, {Radhakrishnan}, {Doelman},
  {van Kooten}  \& {Bos}}{{Por} et~al.}{2018}]{2018SPIE10703E..42P}
{Por} E.~H.,  {Haffert} S.~Y.,  {Radhakrishnan} V.~M.,  {Doelman} D.~S.,  {van
  Kooten} M.,   {Bos} S.~P.,  2018, in {Close} L.~M.,  {Schreiber} L.,
  {Schmidt} D.,  eds,  Society of Photo-Optical Instrumentation Engineers
  (SPIE) Conference Series Vol. 10703, Adaptive Optics Systems VI. p. 1070342,
  \mn@doi{10.1117/12.2314407}

\bibitem[\protect\citeauthoryear{{Rabien} et~al.,}{{Rabien}
  et~al.}{2019}]{2019A&A...621A...4R}
{Rabien} S.,  et~al., 2019, \mn@doi [\aap] {10.1051/0004-6361/201833716}, \href
  {https://ui.adsabs.harvard.edu/abs/2019A&A...621A...4R} {621, A4}

\bibitem[\protect\citeauthoryear{{Radigan}, {Lafreni{\`e}re}, {Jayawardhana}
  \& {Artigau}}{{Radigan} et~al.}{2014}]{2014ApJ...793...75R}
{Radigan} J.,  {Lafreni{\`e}re} D.,  {Jayawardhana} R.,   {Artigau} E.,  2014,
  \mn@doi [\apj] {10.1088/0004-637X/793/2/75}, \href
  {https://ui.adsabs.harvard.edu/abs/2014ApJ...793...75R} {793, 75}

\bibitem[\protect\citeauthoryear{Reback et~al.,}{Reback
  et~al.}{2022}]{jeff_reback_2022_6408044}
Reback J.,  et~al., 2022, pandas-dev/pandas: Pandas 1.4.2,
  \mn@doi{10.5281/zenodo.6408044}, \url
  {https://doi.org/10.5281/zenodo.6408044}

\bibitem[\protect\citeauthoryear{{Ruane} et~al.,}{{Ruane}
  et~al.}{2018}]{2018SPIE10698E..2SR}
{Ruane} G.,  et~al., 2018, in {Lystrup} M.,  {MacEwen} H.~A.,  {Fazio} G.~G.,
  {Batalha} N.,  {Siegler} N.,   {Tong} E.~C.,  eds,  Society of Photo-Optical
  Instrumentation Engineers (SPIE) Conference Series Vol. 10698, Space
  Telescopes and Instrumentation 2018: Optical, Infrared, and Millimeter Wave.
  p. 106982S (\mn@eprint {arXiv} {1807.07042}), \mn@doi{10.1117/12.2312948}

\bibitem[\protect\citeauthoryear{{Sauvage}, {Fusco}, {Rousset}  \&
  {Petit}}{{Sauvage} et~al.}{2007}]{2007JOSAA..24.2334S}
{Sauvage} J.-F.,  {Fusco} T.,  {Rousset} G.,   {Petit} C.,  2007, \mn@doi
  [Journal of the Optical Society of America A] {10.1364/JOSAA.24.002334},
  \href {https://ui.adsabs.harvard.edu/abs/2007JOSAA..24.2334S} {24, 2334}

\bibitem[\protect\citeauthoryear{{Scargle}}{{Scargle}}{1982}]{1982ApJ...263..835S}
{Scargle} J.~D.,  1982, \mn@doi [\apj] {10.1086/160554}, \href
  {https://ui.adsabs.harvard.edu/abs/1982ApJ...263..835S} {263, 835}

\bibitem[\protect\citeauthoryear{Seabold \& Perktold}{Seabold \&
  Perktold}{2010}]{seabold2010statsmodels}
Seabold S.,  Perktold J.,  2010, in 9th Python in Science Conference.

\bibitem[\protect\citeauthoryear{{Skaf} et~al.,}{{Skaf}
  et~al.}{2022}]{2022A&A...659A.170S}
{Skaf} N.,  et~al., 2022, \mn@doi [\aap] {10.1051/0004-6361/202141514}, \href
  {https://ui.adsabs.harvard.edu/abs/2022A&A...659A.170S} {659, A170}

\bibitem[\protect\citeauthoryear{{Skemer} et~al.,}{{Skemer}
  et~al.}{2014}]{2014SPIE.9148E..0LS}
{Skemer} A.~J.,  et~al., 2014, in {Marchetti} E.,  {Close} L.~M.,   {Vran}
  J.-P.,  eds,  Society of Photo-Optical Instrumentation Engineers (SPIE)
  Conference Series Vol. 9148, Adaptive Optics Systems IV. p. 91480L
  (\mn@eprint {arXiv} {1407.2876}), \mn@doi{10.1117/12.2057277}

\bibitem[\protect\citeauthoryear{{Skemer} et~al.,}{{Skemer}
  et~al.}{2015}]{2015SPIE.9605E..1DS}
{Skemer} A.~J.,  et~al., 2015, in {Shaklan} S.,  ed.,  Society of Photo-Optical
  Instrumentation Engineers (SPIE) Conference Series Vol. 9605, Techniques and
  Instrumentation for Detection of Exoplanets VII. p. 96051D (\mn@eprint
  {arXiv} {1508.06290}), \mn@doi{10.1117/12.2187284}

\bibitem[\protect\citeauthoryear{{Skemer}, {Hinz}, {Stone}, {Skrutskie},
  {Woodward}, {Leisenring}  \& {Briesemeister}}{{Skemer}
  et~al.}{2018}]{2018SPIE10702E..0CS}
{Skemer} A.~J.,  {Hinz} P.,  {Stone} J.,  {Skrutskie} M.,  {Woodward} C.~E.,
  {Leisenring} J.,   {Briesemeister} Z.,  2018, in {Evans} C.~J.,  {Simard} L.,
    {Takami} H.,  eds,  Society of Photo-Optical Instrumentation Engineers
  (SPIE) Conference Series Vol. 10702, Ground-based and Airborne
  Instrumentation for Astronomy VII. p. 107020C (\mn@eprint {arXiv}
  {1808.03301}), \mn@doi{10.1117/12.2314091}

\bibitem[\protect\citeauthoryear{{Skrutskie} et~al.,}{{Skrutskie}
  et~al.}{2010}]{2010SPIE.7735E..3HS}
{Skrutskie} M.~F.,  et~al., 2010, in {McLean} I.~S.,  {Ramsay} S.~K.,
  {Takami} H.,  eds,  Society of Photo-Optical Instrumentation Engineers (SPIE)
  Conference Series Vol. 7735, Ground-based and Airborne Instrumentation for
  Astronomy III. p. 77353H, \mn@doi{10.1117/12.857724}

\bibitem[\protect\citeauthoryear{{Snik}, {Otten}, {Kenworthy}, {Miskiewicz},
  {Escuti}, {Packham}  \& {Codona}}{{Snik} et~al.}{2012}]{2012SPIE.8450E..0MS}
{Snik} F.,  {Otten} G.,  {Kenworthy} M.,  {Miskiewicz} M.,  {Escuti} M.,
  {Packham} C.,   {Codona} J.,  2012, in {Navarro} R.,  {Cunningham} C.~R.,
  {Prieto} E.,  eds,  Society of Photo-Optical Instrumentation Engineers (SPIE)
  Conference Series Vol. 8450, Modern Technologies in Space- and Ground-based
  Telescopes and Instrumentation II. p. 84500M (\mn@eprint {arXiv}
  {1207.2970}), \mn@doi{10.1117/12.926222}

\bibitem[\protect\citeauthoryear{{Stolker}, {Bonse}, {Quanz}, {Amara}, {Cugno},
  {Bohn}  \& {Boehle}}{{Stolker} et~al.}{2019}]{2019A&A...621A..59S}
{Stolker} T.,  {Bonse} M.~J.,  {Quanz} S.~P.,  {Amara} A.,  {Cugno} G.,  {Bohn}
  A.~J.,   {Boehle} A.,  2019, \mn@doi [\aap] {10.1051/0004-6361/201834136},
  \href {https://ui.adsabs.harvard.edu/abs/2019A&A...621A..59S} {621, A59}

\bibitem[\protect\citeauthoryear{{Stone}, {Skemer}, {Hinz}, {Briesemeister},
  {Barman}, {Woodward}, {Skrutskie}  \& {Leisenring}}{{Stone}
  et~al.}{2018}]{2018SPIE10702E..3FS}
{Stone} J.~M.,  {Skemer} A.~J.,  {Hinz} P.,  {Briesemeister} Z.,  {Barman} T.,
  {Woodward} C.~E.,  {Skrutskie} M.,   {Leisenring} J.,  2018, in {Evans}
  C.~J.,  {Simard} L.,   {Takami} H.,  eds,  Society of Photo-Optical
  Instrumentation Engineers (SPIE) Conference Series Vol. 10702, Ground-based
  and Airborne Instrumentation for Astronomy VII. p. 107023F (\mn@eprint
  {arXiv} {1808.02571}), \mn@doi{10.1117/12.2313977}

\bibitem[\protect\citeauthoryear{{Stone} et~al.,}{{Stone}
  et~al.}{2020}]{2020AJ....160..262S}
{Stone} J.~M.,  et~al., 2020, \mn@doi [\aj] {10.3847/1538-3881/abbef3}, \href
  {https://ui.adsabs.harvard.edu/abs/2020AJ....160..262S} {160, 262}

\bibitem[\protect\citeauthoryear{{Stone} et~al.,}{{Stone}
  et~al.}{2022}]{2022SPIE12184E..42S}
{Stone} J.~M.,  et~al., 2022, in {Evans} C.~J.,  {Bryant} J.~J.,   {Motohara}
  K.,  eds,  Society of Photo-Optical Instrumentation Engineers (SPIE)
  Conference Series Vol. 12184, Ground-based and Airborne Instrumentation for
  Astronomy IX. p. 1218442, \mn@doi{10.1117/12.2630308}

\bibitem[\protect\citeauthoryear{{Sutlieff} et~al.,}{{Sutlieff}
  et~al.}{2021}]{2021MNRAS.506.3224S}
{Sutlieff} B.~J.,  et~al., 2021, \mn@doi [\mnras] {10.1093/mnras/stab1893},
  \href {https://ui.adsabs.harvard.edu/abs/2021MNRAS.506.3224S} {506, 3224}

\bibitem[\protect\citeauthoryear{{Sutlieff} et~al.,}{{Sutlieff}
  et~al.}{2023}]{2023MNRAS.520.4235S}
{Sutlieff} B.~J.,  et~al., 2023, \mn@doi [\mnras] {10.1093/mnras/stad249},
  \href {https://ui.adsabs.harvard.edu/abs/2023MNRAS.520.4235S} {520, 4235}

\bibitem[\protect\citeauthoryear{{Sutlieff} et~al.,}{{Sutlieff}
  et~al.}{2024}]{2024MNRAS.531.2168S}
{Sutlieff} B.~J.,  et~al., 2024, \mn@doi [\mnras] {10.1093/mnras/stae1315},
  \href {https://ui.adsabs.harvard.edu/abs/2024MNRAS.531.2168S} {531, 2168}

\bibitem[\protect\citeauthoryear{{Tan} \& {Showman}}{{Tan} \&
  {Showman}}{2019}]{2019ApJ...874..111T}
{Tan} X.,  {Showman} A.~P.,  2019, \mn@doi [\apj] {10.3847/1538-4357/ab0c07},
  \href {https://ui.adsabs.harvard.edu/abs/2019ApJ...874..111T} {874, 111}

\bibitem[\protect\citeauthoryear{{Tannock} et~al.,}{{Tannock}
  et~al.}{2021}]{2021AJ....161..224T}
{Tannock} M.~E.,  et~al., 2021, \mn@doi [\aj] {10.3847/1538-3881/abeb67}, \href
  {https://ui.adsabs.harvard.edu/abs/2021AJ....161..224T} {161, 224}

\bibitem[\protect\citeauthoryear{{Todorov} et~al.,}{{Todorov}
  et~al.}{2019}]{2019A&A...631A.169T}
{Todorov} K.~O.,  et~al., 2019, \mn@doi [\aap] {10.1051/0004-6361/201935364},
  \href {https://ui.adsabs.harvard.edu/abs/2019A&A...631A.169T} {631, A169}

\bibitem[\protect\citeauthoryear{{Tremblin}, {Amundsen}, {Chabrier}, {Baraffe},
  {Drummond}, {Hinkley}, {Mourier}  \& {Venot}}{{Tremblin}
  et~al.}{2016}]{2016ApJ...817L..19T}
{Tremblin} P.,  {Amundsen} D.~S.,  {Chabrier} G.,  {Baraffe} I.,  {Drummond}
  B.,  {Hinkley} S.,  {Mourier} P.,   {Venot} O.,  2016, \mn@doi [\apjl]
  {10.3847/2041-8205/817/2/L19}, \href
  {https://ui.adsabs.harvard.edu/abs/2016ApJ...817L..19T} {817, L19}

\bibitem[\protect\citeauthoryear{{Tremblin}, {Phillips}, {Emery}, {Baraffe},
  {Lew}, {Apai}, {Biller}  \& {Bonnefoy}}{{Tremblin}
  et~al.}{2020}]{2020A&A...643A..23T}
{Tremblin} P.,  {Phillips} M.~W.,  {Emery} A.,  {Baraffe} I.,  {Lew} B.~W.~P.,
  {Apai} D.,  {Biller} B.~A.,   {Bonnefoy} M.,  2020, \mn@doi [\aap]
  {10.1051/0004-6361/202038771}, \href
  {https://ui.adsabs.harvard.edu/abs/2020A&A...643A..23T} {643, A23}

\bibitem[\protect\citeauthoryear{{Vigan} et~al.,}{{Vigan}
  et~al.}{2019}]{2019A&A...629A..11V}
{Vigan} A.,  et~al., 2019, \mn@doi [\aap] {10.1051/0004-6361/201935889}, \href
  {https://ui.adsabs.harvard.edu/abs/2019A&A...629A..11V} {629, A11}

\bibitem[\protect\citeauthoryear{{Vigan} et~al.,}{{Vigan}
  et~al.}{2022}]{2022A&A...660A.140V}
{Vigan} A.,  et~al., 2022, \mn@doi [\aap] {10.1051/0004-6361/202142635}, \href
  {https://ui.adsabs.harvard.edu/abs/2022A&A...660A.140V} {660, A140}

\bibitem[\protect\citeauthoryear{Virtanen et~al.,}{Virtanen
  et~al.}{2020}]{2020SciPy-NMeth}
Virtanen P.,  et~al., 2020, \mn@doi [Nature Methods]
  {10.1038/s41592-019-0686-2}, \href {https://rdcu.be/b08Wh} {17, 261}

\bibitem[\protect\citeauthoryear{{Vos} et~al.,}{{Vos}
  et~al.}{2019}]{2019MNRAS.483..480V}
{Vos} J.~M.,  et~al., 2019, \mn@doi [\mnras] {10.1093/mnras/sty3123}, \href
  {https://ui.adsabs.harvard.edu/abs/2019MNRAS.483..480V} {483, 480}

\bibitem[\protect\citeauthoryear{{Vos}, {Faherty}, {Gagn{\'e}}, {Marley},
  {Metchev}, {Gizis}, {Rice}  \& {Cruz}}{{Vos}
  et~al.}{2022}]{2022ApJ...924...68V}
{Vos} J.~M.,  {Faherty} J.~K.,  {Gagn{\'e}} J.,  {Marley} M.,  {Metchev} S.,
  {Gizis} J.,  {Rice} E.~L.,   {Cruz} K.,  2022, \mn@doi [\apj]
  {10.3847/1538-4357/ac4502}, \href
  {https://ui.adsabs.harvard.edu/abs/2022ApJ...924...68V} {924, 68}

\bibitem[\protect\citeauthoryear{{Vos} et~al.,}{{Vos}
  et~al.}{2023}]{2023ApJ...944..138V}
{Vos} J.~M.,  et~al., 2023, \mn@doi [\apj] {10.3847/1538-4357/acab58}, \href
  {https://ui.adsabs.harvard.edu/abs/2023ApJ...944..138V} {944, 138}

\bibitem[\protect\citeauthoryear{{Wang}, {Kulikauskas}  \& {Blunt}}{{Wang}
  et~al.}{2021}]{2021ascl.soft01003W}
{Wang} J.~J.,  {Kulikauskas} M.,   {Blunt} S.,  2021, {whereistheplanet:
  Predicting positions of directly imaged companions}, Astrophysics Source Code
  Library, record ascl:2101.003 (\mn@eprint {ascl} {2101.003})

\bibitem[\protect\citeauthoryear{{Wang} et~al.,}{{Wang}
  et~al.}{2022}]{2022AJ....164..143W}
{Wang} J.~J.,  et~al., 2022, \mn@doi [\aj] {10.3847/1538-3881/ac8984}, \href
  {https://ui.adsabs.harvard.edu/abs/2022AJ....164..143W} {164, 143}

\bibitem[\protect\citeauthoryear{{Wenger} et~al.,}{{Wenger}
  et~al.}{2000}]{2000A&AS..143....9W}
{Wenger} M.,  et~al., 2000, \mn@doi [\aaps] {10.1051/aas:2000332}, \href
  {https://ui.adsabs.harvard.edu/abs/2000A&AS..143....9W} {143, 9}

\bibitem[\protect\citeauthoryear{{Wilson}, {Rajan}  \& {Patience}}{{Wilson}
  et~al.}{2014}]{2014A&A...566A.111W}
{Wilson} P.~A.,  {Rajan} A.,   {Patience} J.,  2014, \mn@doi [\aap]
  {10.1051/0004-6361/201322995}, \href
  {https://ui.adsabs.harvard.edu/abs/2014A&A...566A.111W} {566, A111}

\bibitem[\protect\citeauthoryear{{Yang} et~al.,}{{Yang}
  et~al.}{2016}]{2016ApJ...826....8Y}
{Yang} H.,  et~al., 2016, \mn@doi [\apj] {10.3847/0004-637X/826/1/8}, \href
  {https://ui.adsabs.harvard.edu/abs/2016ApJ...826....8Y} {826, 8}

\bibitem[\protect\citeauthoryear{{Zernike}}{{Zernike}}{1934}]{1934Phy.....1..689Z}
{Zernike} v.~F.,  1934, \mn@doi [Physica] {10.1016/S0031-8914(34)80259-5},
  \href {https://ui.adsabs.harvard.edu/abs/1934Phy.....1..689Z} {1, 689}

\bibitem[\protect\citeauthoryear{{Zhou}, {Apai}, {Schneider}, {Marley}  \&
  {Showman}}{{Zhou} et~al.}{2016}]{2016ApJ...818..176Z}
{Zhou} Y.,  {Apai} D.,  {Schneider} G.~H.,  {Marley} M.~S.,   {Showman} A.~P.,
  2016, \mn@doi [\apj] {10.3847/0004-637X/818/2/176}, \href
  {https://ui.adsabs.harvard.edu/abs/2016ApJ...818..176Z} {818, 176}

\bibitem[\protect\citeauthoryear{{Zhou}, {Bowler}, {Morley}, {Apai}, {Kataria},
  {Bryan}  \& {Benneke}}{{Zhou} et~al.}{2020}]{2020AJ....160...77Z}
{Zhou} Y.,  {Bowler} B.~P.,  {Morley} C.~V.,  {Apai} D.,  {Kataria} T.,
  {Bryan} M.~L.,   {Benneke} B.,  2020, \mn@doi [\aj]
  {10.3847/1538-3881/ab9e04}, \href
  {https://ui.adsabs.harvard.edu/abs/2020AJ....160...77Z} {160, 77}

\bibitem[\protect\citeauthoryear{{Zhou}, {Bowler}, {Apai}, {Kataria}, {Morley},
  {Bryan}, {Skemer}  \& {Benneke}}{{Zhou} et~al.}{2022}]{2022AJ....164..239Z}
{Zhou} Y.,  {Bowler} B.~P.,  {Apai} D.,  {Kataria} T.,  {Morley} C.~V.,
  {Bryan} M.~L.,  {Skemer} A.~J.,   {Benneke} B.,  2022, \mn@doi [\aj]
  {10.3847/1538-3881/ac9905}, \href
  {https://ui.adsabs.harvard.edu/abs/2022AJ....164..239Z} {164, 239}

\bibitem[\protect\citeauthoryear{{de Mooij}, {de Kok}, {Nefs}  \&
  {Snellen}}{{de Mooij} et~al.}{2011}]{2011A&A...528A..49D}
{de Mooij} E.~J.~W.,  {de Kok} R.~J.,  {Nefs} S.~V.,   {Snellen} I.~A.~G.,
  2011, \mn@doi [\aap] {10.1051/0004-6361/201016142}, \href
  {https://ui.adsabs.harvard.edu/abs/2011A&A...528A..49D} {528, A49}

\bibitem[\protect\citeauthoryear{{van Kooten}, {Doelman}  \& {Kenworthy}}{{van
  Kooten} et~al.}{2020}]{2020A&A...636A..81V}
{van Kooten} M.~A.~M.,  {Doelman} N.,   {Kenworthy} M.,  2020, \mn@doi [\aap]
  {10.1051/0004-6361/201937076}, \href
  {https://ui.adsabs.harvard.edu/abs/2020A&A...636A..81V} {636, A81}

\bibitem[\protect\citeauthoryear{{van Kooten}, {Jensen-Clem}, {Cetre},
  {Ragland}, {Bond}, {Fowler}  \& {Wizinowich}}{{van Kooten}
  et~al.}{2022}]{2022JATIS...8b9006V}
{van Kooten} M. A.~M.,  {Jensen-Clem} R.,  {Cetre} S.,  {Ragland} S.,  {Bond}
  C.~Z.,  {Fowler} J.,   {Wizinowich} P.,  2022, \mn@doi [Journal of
  Astronomical Telescopes, Instruments, and Systems]
  {10.1117/1.JATIS.8.2.029006}, \href
  {https://ui.adsabs.harvard.edu/abs/2022JATIS...8b9006V} {8, 029006}

\makeatother
\end{thebibliography}





\appendix



\bsp	
\label{lastpage}
\end{document}